| Title | **Seed-packed dielectric barrier device for plasma agriculture: understanding its electrical properties through an equivalent electrical model** |
|---|---|
| **Authors** | F. Judée, T. Dufour |
| **Affiliations** | LPP, Sorbonne Université Univ. Paris 6, CNRS, Ecole Polytech., Univ. Paris-Sud, Observatoire de Paris, Université Paris-Saclay,10 PSL Research University, 4 Place Jussieu, 75252 Paris, France. E-mail: thierry.dufour@sorbonne-universite.fr |
| **Ref.** | J. Appl. Phys., Vol. 128, Issue 4, 2020 |
| **DOI** | https://dx.doi.org/10.1063/1.5139889 |
| **Abstract** | Seeds have been packed in a dielectric barrier device where cold atmospheric plasma has been generated to improve their germinative properties. A special attention has been paid on understanding the resulting plasma electrical properties through an equivalent electrical model whose experimental validity has been demonstrated here. In this model, the interelectrode gap is subdivided into 4 types of elementary domains, according to whether they contain electric charges (or not) and according to their type of medium (gas, seed or insulator). The model enables to study the influence of seeds on the plasma electrical properties by measuring and deducing several parameters (charge per filament, gas capacitance, plasma power, …) either in no-bed configuration (i.e. no seed in the reactor) or in packed-bed configuration (seeds in the reactor). In that second case, we have investigated how seeds can influence the plasma electrical parameters considering six specimens of seeds (beans, radishes, corianders, lentils, sunflowers and corns). The influence of molecular oxygen (0-100 sccm) mixed with a continuous flow rate of helium (2 slm) is also investigated, especially through filaments breakdown voltages, charge per filament and plasma power. It is demonstrated that such bed-packing drives to an increase in the gas capacitance ($\xi_{OFF}$), to a decrease in the $\beta^-$ parameter and to variations of the filaments' breakdown voltages in a seed-dependent manner. Finally, we show how the equivalent electrical model can be used to assess the total volume of the contact points, the capacitance of the seeds in the packed-bed configuration and we demonstrate that germinative effects can be induced by plasma on four of the six agronomical specimens. |

# I. Introduction

## I.A. Packed-bed DBD reactors

### I.A.1. Electrical parameters of PB-DBD

Packed-bed dielectric barrier devices (PB-DBD) correspond to a type of plasma source where the interelectrode gap is filled with dielectric or ferroelectric pellets (or beads) usually utilized as catalysts and forming a network through which the ionized gas can flow [1]. The pellets can change the behavior of the micro-discharges through stronger electric field [2] while the plasma-active species can modify the catalyst properties of the pellets, for example by decreasing their activation energy in the case of silver-alumina pellets [3]. Understanding the plasma-pellet interaction is a subject of major interest, especially on the point of applicative view since it is directly connected to technological issues like rising energy efficiency or improving product selectivity.

Despite the difficulty of using optical diagnosis to characterize plasma in PB-DBD configuration, Mujahid et al. successfully utilized phase and space-resolved spectroscopy to show the existence of three mechanisms sustaining plasma: filamentary discharge in the volume (between the center of dielectric structures and the opposite electrode), micro-discharges at the contact points and surface ionization waves over the dielectric surface [4]. They showed that plasma was essentially generated at the contact points between the dielectric structures where an intense electric field is present. Zhang et al. managed to assess this electric field strengthening using a 2D particle-in-cell/Monte Carlo collision model. A value as high as $4.10^8$ V/m was assessed at the contact points while a value as low as $2.10^7$ V/m was obtained in the same DBD without pellets [5]. More recently, Bogaerts et al. explained with a fluid model to which extent electric field increases with pellet dielectric constant and how this effect is even more pronounced in microgap (500 µm) than in mm-gap (4.5 mm) [6]. Besides, it has been shown that this electric field strengthening is linked to an enhancement of electron density and electron energy at the necks between ferroelectric pellets. Gómez-Ramírez et al. evidenced an increase in electron density from $2.10^{14}$ m$^{-3}$ to $3.10^{15}$ m$^{-3}$ on a range of frequency comprised between 500 Hz and 5000 Hz [7]. Electric field strengthening can also drive to localized thermal effects commonly called hotspots and which result either from surface irregularities and/or from the utilization of RF or microwave plasmas [8]. Owing to energetic ions, electrons and electronically excited species impinging on catalyst surface, temperature gradients of several hundreds of °C can locally appear [9].

The sizing of the device plays an important role on the electrical parameters. As demonstrated by Uytdenhouwen et al., lengthening the gap from 2.5mm to 4.7mm makes the discharge less uniform over the entire reaction volume, leading to a decrease in the gap capacitance as well as in the dissipated electrical power [10]. Several characteristics of the pellets (or beads) can also directly impact on the plasma electrical power. Van Laer et al. recommended to use packing beads with diameters of at least one third of the gap size [11] or choosing pellets with higher dielectric permittivity to significantly enhance the breakdown voltages, the plasma density and the current profiles [12]. The aspect ratio of the pellets can also raise local electric field around the contact points between the pellets, driving to a much higher number of micro-discharges at these locations [13]. Finally, it is well known that pellets with larger dielectric constants are more effectively polarized since the electric field is enhanced, driving to an increase in the discharge intensity and electron density [14]. In the case of a PB-DBD in air at atmospheric pressure, 2D particle-in cell/Monte Carlo simulations gave electron density values as low as $7.10^{22}$ $m^{-3}$ for $\varepsilon_r$=9 but higher than $10^{24}$ $m^{-3}$ for $\varepsilon_r$=22 [5]. At the light of these works, pellets capacitance appears as a key parameter to promote plasma catalytic reactions at contact points.

The capacitance of the interelectrode gap, and in particular of the pellets, can be determined using an LCRmeter when the plasma is switched off. An alternative to measure this parameter when the plasma is in operation is based on the use of an equivalent electrical model and experimental data from Lissajous diagrams, as proposed in this article. Plotting a Lissajous diagram is a simple and powerful method to determine its most relevant electrical parameters, e.g. gap capacitance, breakdown and burn voltages, charge per filament, electric energy...

#### I.A.2. From chemical engineering to agriculture applications

For at least two decades, the plasma-pellets interactions have been studied in various packed-bed plasma reactors, especially in dielectric barrier devices (DBD) to cover a wide range of applications. At the early 2000, packed-bed plasma reactors were designed for the destruction or synthesis of chemicals in aqueous phase, e.g. the synthesis of dinitrogen pentoxide ($N_2O_5$) from $NO_2$ [15]. They were also investigated for the destruction of toxic permanent molecules, especially volatile organic compounds such as dichloromethane whose destruction could be achieved in an argon-oxygen discharge with barium titanate beads [16]. Under optimal conditions, Fitzsimmons et al. obtained dichloromethane destructions rates as high as 20% although CO, $COCl_2$, HCl, and $Cl_2$ end-products remained always present. PB-DBD reactors were also assessed for the conversion of many other VOCs (Volatile Organic Compounds), especially methane, xylene, styrene, toluene and benzene [17]. After 2010, they were widely investigated for gaseous pollutant removal, especially the conversion and the valorization of $CO_2$ using various catalysts like Ni–zeolite, $BaTiO_3$ or $ZrO_2$ pellets [18], [19] [20].

Today, plasma-pellets interactions in PB-DBD reactors can be envisioned under a new prism to respond issues different from those dealing with gas valorization. Instead of using pellets to catalyze gaseous chemical reactions, plasma can be used to intentionally change the properties of the pellets. This paradigm shift seems to be particularly interesting for agriculture applications: seeds can be substituted to material pellets with the aim to improve their germinative properties.

Owing to their specific sizes, aspect ratios, moisture content, starch, protein and fat reserves, seeds present electrical conductivity and permittivity parameters, enabling their modeling by resistor-conductor dipoles [21]. Since the values of these parameters can significantly differ from one type of seed to another one [22], their packing into a DBD reactor is expected to change plasma through its electrical (and therefore physico-chemical) properties, even if identical setpoint parameters are imposed (gas flow rate, power, gap distance, etc.). One of the key parameters, which is also an unknown, is the capacitance of the seeds' bed. This parameter could drastically change electric field distribution from one type of seed to the other, as well as the resulting parameters like electron temperature, electron density, charge per filament and plasma power. Being able to evaluate these parameters and control their values represents a major asset since they govern the chemical reactions leading to the production of reactive species near the contact points.

## I.B. Relevance of PB-DBD to meet agriculture stakes

#### I.B.1. Positioning of PB-DBD to other dry/wet plasma processes

Although several types of dry plasma processes have already been employed to treat seeds, to the best of the authors' knowledge, no experimental work has been published on PB-DBD reactors utilized in that respect. Among the current technological dry plasma processes, Meng et al. have improved the germination of wheat seeds placed on the grounded electrode of a DBD reactor without - however - any packed bed [23]. Germination potential was significantly increased by 24% after 4 min of treatment. Corona plasma jets have also been utilized to enhance the germination rate of rape seeds while drastically reducing their micro-organisms loads, in particular aerobic bacteria [24]. Another alternative relies on RF plasma processes operating in air with the ability to treat cereal and legume seeds without any bed configuration. There, the two copper electrodes operated in a vacuum chamber and were externally cooled by water [25]. Finally, low pressure microwave plasma treatments were successfully applied on seeds of Lamb's Quarters to increase their germination rate by a factor 3 without, once again, any packed bed configuration [26].

The diversity of these configurations must not stray from the specific needs of the agricultural sector where the concepts of agriculture yield and industrial productivity must prevail. In that regard, the aforementioned dry plasma processes present a major asset: the seeds can be stored on a certain period without any risk of germination. On the contrary, the wet plasma approaches have a common limitation: plasma-activated water cannot be stored over long periods owing to reactive species lifetimes [27]. Although plasma-activated liquids have already demonstrated spectacular biological effects on seedlings [28], they must be utilized in a reasonable delay to prevent any ageing effects potentially emphasized by environmental factors (sunlight, temperature, …). For these reasons, one should envision dry and wet plasma processes as complementary approaches to effectively meet the challenges of contemporary agriculture.

#### I.B.2. Biological and non-biological effects induced by dry plasma processes on seeds

Seeds germination (but also vigor) is controlled by two main phytohormones: (i) the abscisic acid which is a positive regulator of dormancy induction and a negative regulator of germination and (ii) the gibberellic acid which releases dormancy, promotes germination and counteracts the effects of abscisic acid [29]. It has already been demonstrated that the dry treatment of sunflower seeds using DBD has efficiently induced changes in phytohormone content: the abscisic acid amount could be reduced by 40% while the levels of gibberellic acid could increase up to 43% [30]. The triggering and the production rates of these hormones depend on exogenous factors, especially reactive oxygen/nitrogen species (RONS) which behave as signal molecules [31]. Other works have demonstrated how the germination and enzymatic activity of mung beans can be enhanced using an RF low pressure dry plasma process [32]. Compared to the untreated samples, amylases and proteases could be increased by +25.2% and +36.7% respectively. On the contrary, the trypsin inhibitor activity could be reduced by 39.23% compared to the control after 24h of germination. Until now, even if the reactive oxygen and nitrogen species are considered to be responsible for the aforementioned biological effects, there is no consensus around one reactive plasma species responsible for these effects.

More recently, short plasma treatments (40 s) increased biomass in both shoot and roots by 57%, while longer plasma exposures (80 s) delayed growth and reduced it by 48% [33]. The authors showed how plasma-priming can up-regulate the expression of genes involved in the productions of secondary metabolites, like cannabinoids, and subsequently involve the modification of genes expressions at the transcriptional level.

Dry plasma processes can also induce non-biological modifications of seeds, especially surface modification of their teguments, which result in a modification of their wetting properties. Sera et al have listed seeds with plasma-induced surface modifications and explained how surface functionalization and surface etching can promote seeds germination in a seed-dependent manner [34]. In the works of Bormashenko et al., coatings of lentils, beans and wheat were partially oxidized after RF plasma exposure [35]. TOF-SIMS mass spectrometry evidenced a strong incorporation of nitrogen-containing groups, considered as promoters of germination.

Finally, dry plasma processes can be utilized in agriculture to answer decontamination issues. The production of RONS can indeed eliminate various pathogenic agents, in particular the fungi and the bactericidal loads present on/in the seeds and seedlings. Schnabel et al. have demonstrated how DBD plasma or microwave plasma processed air could decontaminate seeds of *Brassica napus* from endospores of *Bacillus atrophaeus* [36]. Reduction rates comprised between 0.5 and 5.2 log were achieved after 15 min of plasma exposure, without affecting seeds viability. Atmospheric pressure DBD has also been successfully utilized to decontaminate sprout seeds of alfalfa. After 15 min of treatment, inactivation efficiencies as high as 2.04 log and 2.47 log were obtained on natural microbiota and artificially-applied *E. coli* respectively [37]. Similar levels of decontamination were also reached on rapeseed (Brassica napus L.) seeds using a corona discharge plasma jet. After 3 min of plasma exposure, 2.2 and 2.0 log reductions in total aerobic bacteria and E. coli were obtained [38]. So far, no high microbial reduction has been demonstrated but plasma treatments could be combined with other antimicrobial methods like chemicals and lead to synergistic and stronger effects.

# II. Experimental setup and equivalent electrical model of the DBD reactor

## II.A. Experimental setup

In this experimental work, the seeds are treated by plasma in a dielectric barrier device (DBD) composed of a rectangular quartz tube ($\varepsilon_r = 4.5$), 300 mm long, 1.5 mm wall thick, 30 mm internal length and 10 mm internal width, as depicted in Figure 1a. Two plate electrodes are placed in regard along the tube, hence designing an interelectrode region whose parallelepipedal volume is defined by an area of 30 cm² and an interelectrode distance of 11 mm. Similarly, the gap region is defined as a parallelepiped volume $V_{gap}$ (area = 3×10 cm², gap = 10 mm) where gas is ionized with/without seeds. The DBD reactor operates either without seeds (no bed configuration, NB) or with seeds (packed bed configuration, PB). The inner electrode of the DBD is grounded while the outer electrode is polarized to a sine-wave high voltage generated using a function generator (Centrad, GF 467 AF) coupled with an audio amplifier (Crest Audio, CC 5500 W). For all the experiments carried out in this article, the supplied voltage is fixed to 8 kV in magnitude at a frequency of 600 Hz. Two electrical signals are monitored: (i) the output signal of the generator $V_{gene}$ and (ii) the overall electrical charge $Q_{reactor}$ transferred to the reactor and measured by assessing the voltage $V_m$ across the capacitor $C_m$ = 100 pF, as sketched in Figure 1a. $V_{gene}$ is measured using a Tektronix P6015A 1000:1 high voltage probe while $V_m$ is measurer with a Teledyne LeCroy PPE 20 kV 1000:1 high voltage probe. Finally, high precision capacitance measurements were performed using an LCRmeter (HM-8118 model from Rohde & Schwarz).

Q-V diagrams, commonly called Lissajous curves, consist into plotting the charge transferred to the reactor ($Q_{reactor}$) as a function of the voltage applied along its interelectrode distance ($V_{reactor}$) according to equation {1}. Plotting these $(Q\text{-}V)_{reactor}$ diagrams are an easy-to-handle method to measure electrical power of DBD reactors as well as to infer additional discharge properties [39], [40]. In this article, all the Lissajous diagrams are plotted considering charge and voltage over a single period (or cycle).

The DBD reactor is seed-packed in ambient air and supplied in helium gas at 2 slm with/without molecular oxygen (0-100 sccm). All plasma treatments are achieved according to the same following protocol: (i) seed-packing of the DBD reactor, (ii) flushing of the gas line and of the DBD reactor with helium gas during two minutes to reduce as significantly as possible residual background from ambient air, (iii) plasma ignition. No gas purification process is achieved (e.g. PB-DBD placed in an enclosure where primary vacuum in achieved and followed by a return to atmospheric pressure under helium gas). On the point of seed companies view, this additional step would represent investment costs, human interventions and additional delays that would make plasma processes uncompetitive with already marketed alternatives. The background of ambient air is not an issue as long as it is finely characterized. Mass spectrometry measurements, reported in Table 1, indicate the chemical composition of the PB-DBD supplied with helium at 2 slm.

| Species | Peak intensities (arb. units) |
|---|---|
| He | $1.7\times10^6$ |
| $N_2$ | $2.3\times10^4$ |
| $O_2$ | $6.2\times10^3$ |
| $H_2O$ | $5.2\times10^3$ |
| $CO_2$ | $2.0\times10^3$ |

***Table 1. Mass spectrum peak intensities measured in the DBD reactor supplied with pure helium gas***

Photographs of the DBD with/without seeds have been taken with a Reflex Nikon camera (Model D610). The values of exposure time and aperture were set at 1 s and 7.1 respectively to clearly observe the low-emissive micro-discharges. Since roughly 20 micro-discharges follow one another per period (T = 1/600Hz = 1.6 ms), the photographs of the plasma represent a statistical accumulation of micro-discharges over an observation window of 1s. The uniform appearance of the light emitted by plasma indicates a random distribution of the micro-discharges throughout the interelectrode gap.

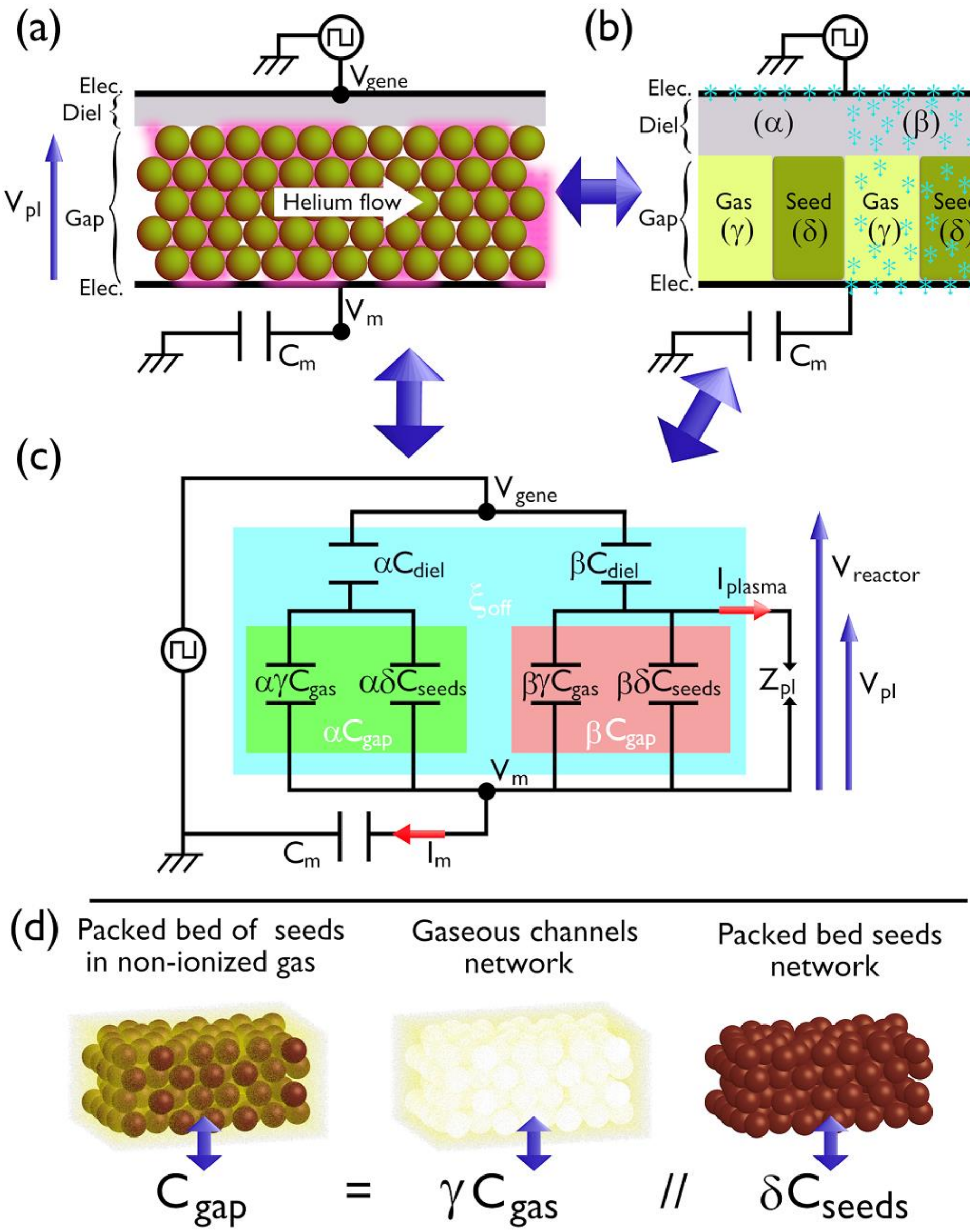

***Fig. 1 (a) Sketch of the experimental setup in cross section view, (b) Synoptic diagram modelling the α, β γ, δ parameters of an elementary packed-bed DBD reactor, (c) equivalent electrical circuit of the DBD reactor that can operate either without seeds (no-bed configuration, NB) or with seeds (packed bed configuration, PB), (d) Tridimensional sketch illustrating how seeds and plasma can be modeled as two parallel capacitive networks.***

## II.B. Equivalent electrical model

The routinely-used equivalent electrical model of a DBD reactor is a dielectric capacitance ($C_{diel}$) placed in series with two dipoles in parallel: a gap capacitance ($C_{gap}$) and a plasma discharge component (represented either as a variable impedance or as a current generator) [41]. If this model is validated for devices where a glow-like plasma is formed between planar electrodes, it appears too simplistic for DBD configurations where the transfer of electrical charges results from a distribution of micro-discharges that is no more random in space and time. Many reasons can explain this non-random character, e.g. the presence of materials in the gap (microparticles, pellets, polymer sample, etc.), the surface state, the non-parallelism of the electrodes, etc.

A more sophisticated equivalent electrical model of DBD reactor has been suggested in Peeters' works [42]. In this model, the author considers that the gap volume within the interelectrode region is divisible into two types of domains: those containing electrical charges and those without any charge. We propose to implement this model by considering the existence of gaseous domains and materials/seeds domains as it is encountered in PB-DBD configurations. This model relies on the following specifications:

(i) The interelectrode region is subdivided in two domains: the dielectric barrier where electrical charges cannot move at macroscopic scale and the gap volume where the electrical charges can move through a distribution of micro-discharges that does not necessarily fill the entire gap volume, potentially leaving some dead volumes, as represented in Figure 1a. Indeed, the distribution of micro-discharges impinging on the electrode (or dielectric barrier) surface cover an area which does not exactly correspond to the full electrode (or dielectric) geometrical area.

(ii) Each of the two previous domains can be subdivided into two types of smaller volumetric cells: those where no electrical charge is present and whose total volume is labelled V' and those which contain electrical charges and represent a total volume V", as sketched in Figure 1b. The equation {2} provides the adimensional numbers associated with V' and V": α and β respectively. The charge conservation is expressed by equation {3} and must always be satisfied.

(iii) The equivalent electrical circuit of a seed can be represented by a capacitor in parallel with a resistor, where capacitance and resistance values depend on excitation frequency [43]. At low frequency (typically less than 1 kHz), the values of resistance remain quite elevated (higher than several 100 kΩ) with capacitance values of few 10-100 pF [44]. In consequence, and since our operating frequency is only 600Hz, the seed equivalent electrical circuit can be reduced to a single capacitor, as achieved here in this work.

(iv) Regardless the existence (or not) of electrical charges, the gap volume can be subdivided into elementary volumetric cells containing either gas or seeds. The sum of all the volumetric cells containing gas is labelled $V_{gas}$ while the total volume occupied by the seeds is called $V_{seeds}$. In equation {4}, the γ and δ parameters report the volumetric fractions of gas and seeds respectively. Since the gap volume $V_{gap}$ is equal to the sum of $V_{gas}$ and $V_{seeds}$, it can be nondimensionalized as expressed in equation {5}.

(v) The capacitance of the dielectric barrier ($C_{diel}$) is numerically determined considering its geometrical shape, relative permittivity ($\varepsilon_r$), thickness ($e_{diel}$) and area ($S_{diel}$), as expressed in equation {6}.

(vi) In the gap volume, all the seeds are in contact with each other, forming a unified dielectric network. Between these seeds exist interstitial spaces that also communicate with each other, forming a parallel network of gaseous channels as depicted in Figure 1d. In the gap volume, the two networks - dielectric and gas - are entangled in one another and can therefore be likened to capacitors in parallel, as expressed in equation {7} where $C_{gas}$ is the capacitance of the non-ionized gas and $C_{seeds}$ is the capacitance of the seeds.

(vii) When plasma is off, the capacitance of the DBD reactor is labelled $\xi_{OFF}$ and, according to formula {8}, can be expressed as the capacitance of the dielectric barrier in series with the capacitance of the gap volume. When the plasma is ignited, the $\xi_{ON}$ capacitance can only include capacitive domains. For this reason, the $\beta C_{gap}$ component stands for the charged domain of the gap volume and cannot be considered, conversely to $\alpha C_{gap}$. As a result, $\xi_{ON}$ can be modelled either as $\alpha\xi_{OFF}$ in parallel with $\beta C_{diel}$ as given in {9} or as $C_{diel}$ in series with $\alpha C_{gap}$ as given in {10}.

(viii) As shown in Figure 2a, the values of $\xi_{OFF}$ and $\xi_{ON}$ can be assessed by plotting the $(Q\text{-}V)_{reactor}$ diagram where $V_{reactor}$ and $Q_{reactor}$ are measured using equations {1} and {11} respectively. The diagram presents a parallelogram shape whose low-slope and high-slope sides correspond to $\xi_{OFF}$ and $\xi_{ON}$ respectively. Depending on the negative or positive half cycle of the applied voltage, it is possible to distinguish $\xi^-_{OFF}$, $\xi^+_{OFF}$, $\xi^-_{ON}$ and $\xi^+_{ON}$ values.

(ix) The α and β parameters can be estimated by solving the system of equations {3} & {9}, with the possibility to distinguish (or not) $\xi^-$ and $\xi^+$ parameters, and in turn $\alpha^-$ and $\alpha^+$ parameters, as well as $\beta^-$ and $\beta^+$ parameters

(x) To distinguish the electrical power dissipated in the interelectrode region ($P_{reactor}$) from the electrical power dissipated in the plasma ($P_{plasma}$), two Lissajous diagrams can be plotted: $(Q\text{-}V)_{reactor}$ and $(Q\text{-}V)_{plasma}$ represented in Figure 2a and 2b respectively. In that second case, the plasma transferred charge ($Q_{plasma}$), plasma voltage ($V_{plasma}$) and electrical plasma power ($P_{plasma}$) are estimated with equations{12}, {13} and {14} respectively.

(xi) Finally, electrical parameters like the transferred charge per filament ($Q_f$), the residual charge ($Q_{res}$) and the breakdown voltages associated with the micro-discharges ($V_{br}$) can be extracted from the (Q-V) diagrams reported in Figure 2. For each of these electrical parameters, one can distinguishes positive and negative values (for example $V^-_{br}$ and $V^+_{br}$ ) depending on negative and positive half-cycles of the supplied voltage. Each charge per filament (i.e. the electrical charge transferred from one electrode to the other) is represented by a segment of negative slope, as shown in Figure 2b. On the parallelogram figure, it is labelled $Q_f^-$ on the left side and $Q_f^+$ on the right side.

$$V_{reactor} = V_{gene} - V_m \; \{1\}$$

$$\alpha = \frac{V'}{V_{gap}} = \frac{V'}{V' + V"} \; \& \; \beta = \frac{V"}{V_{gap}} = \frac{V"}{V' + V"} \; \{2\}$$

$$\alpha + \beta = 1 \; \{3\}$$

$$\gamma = \frac{V_{gas}}{V_{gap}} = \frac{V_{gas}}{V_{gas} + V_{seeds}} \; \& \; \delta = \frac{V_{seeds}}{V_{gap}} = \frac{V_{seeds}}{V_{gas} + V_{seeds}} \; \{4\}$$

$$\gamma + \delta = 1 \; \{5\}$$

$$C_{diel} = \varepsilon_0 . \varepsilon_r . \frac{S_{diel}}{e_{diel}} \; \{6\}$$

$$C_{gap} = \gamma . C_{gas} + \delta . C_{seeds} \; \{7\}$$

$$\frac{1}{\xi_{OFF}} = \frac{1}{C_{diel}} + \frac{1}{C_{gap}} \; \{8\}$$

$$\xi_{ON} = \alpha . \xi_{OFF} + \beta . C_{diel} \; \{9\}$$

$$\frac{1}{\xi_{ON}} = \frac{1}{C_{diel}} + \frac{1}{\alpha . C_{gap}} \; \{10\}$$

$$Q_{reactor} = C_m . V_{reactor} \; \{11\}$$

$$Q_{plasma}(t) = \left( \frac{C_{diel}}{C_{diel} - \xi_{OFF}} \right) \times [Q_{reactor}(t) - \xi_{OFF} . V_{reactor}(t)] \; \{12\}$$

$$V_{plasma}(t) = \left( 1 + \frac{\alpha}{\beta} \frac{\xi_{OFF}}{C_{ciel}} \right) \times \left[ V_{reactor}(t) - \frac{Q_{reactor}(t)}{\beta * C_{diel}} \right] \; \{13\}$$

$$P_{plasma} = \frac{1}{T} \int_T Q_{plasma} . dV_{plasma} \; \{14\}$$

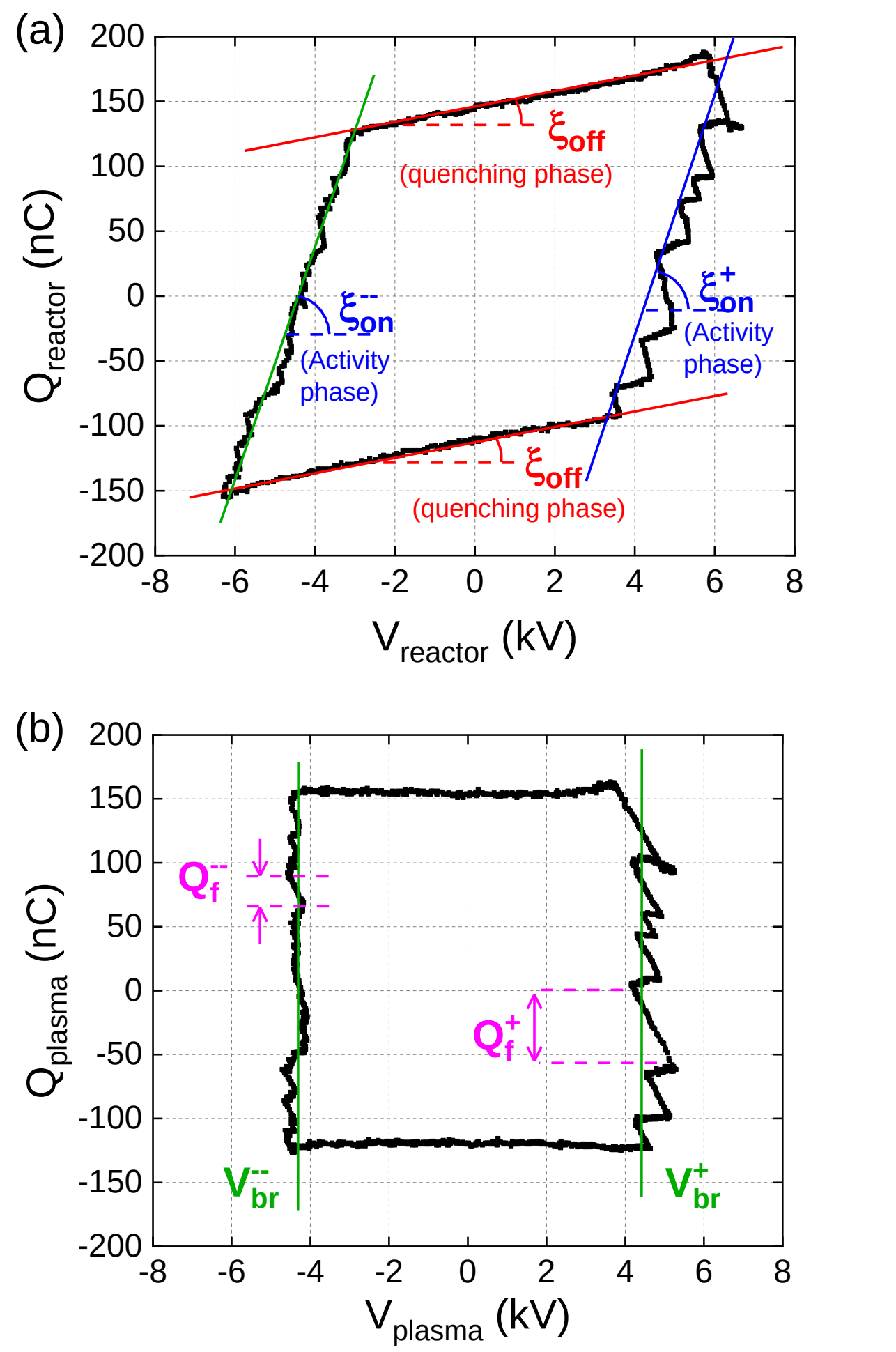

***Fig. 2** Lissajous curves for plasma generated in the NB-DBD configuration(a) (Q-V)$_{reactor}$ diagram of the reactor and (b) (Q-V)$_{plasma}$ diagram of the plasma phase confined into the gap volume of the DBD reactor. All relevant parameters extracted from Q-V diagram are indicated in the figures. Experimental conditions: 600 Hz, Helium 2 slm.*

# III. Results & Discussion

## III.A. Plasma electrical parameters in the NB-DBD reactor

### III.A.1. Experimental validation of the α and β parameters of the equivalent electrical model

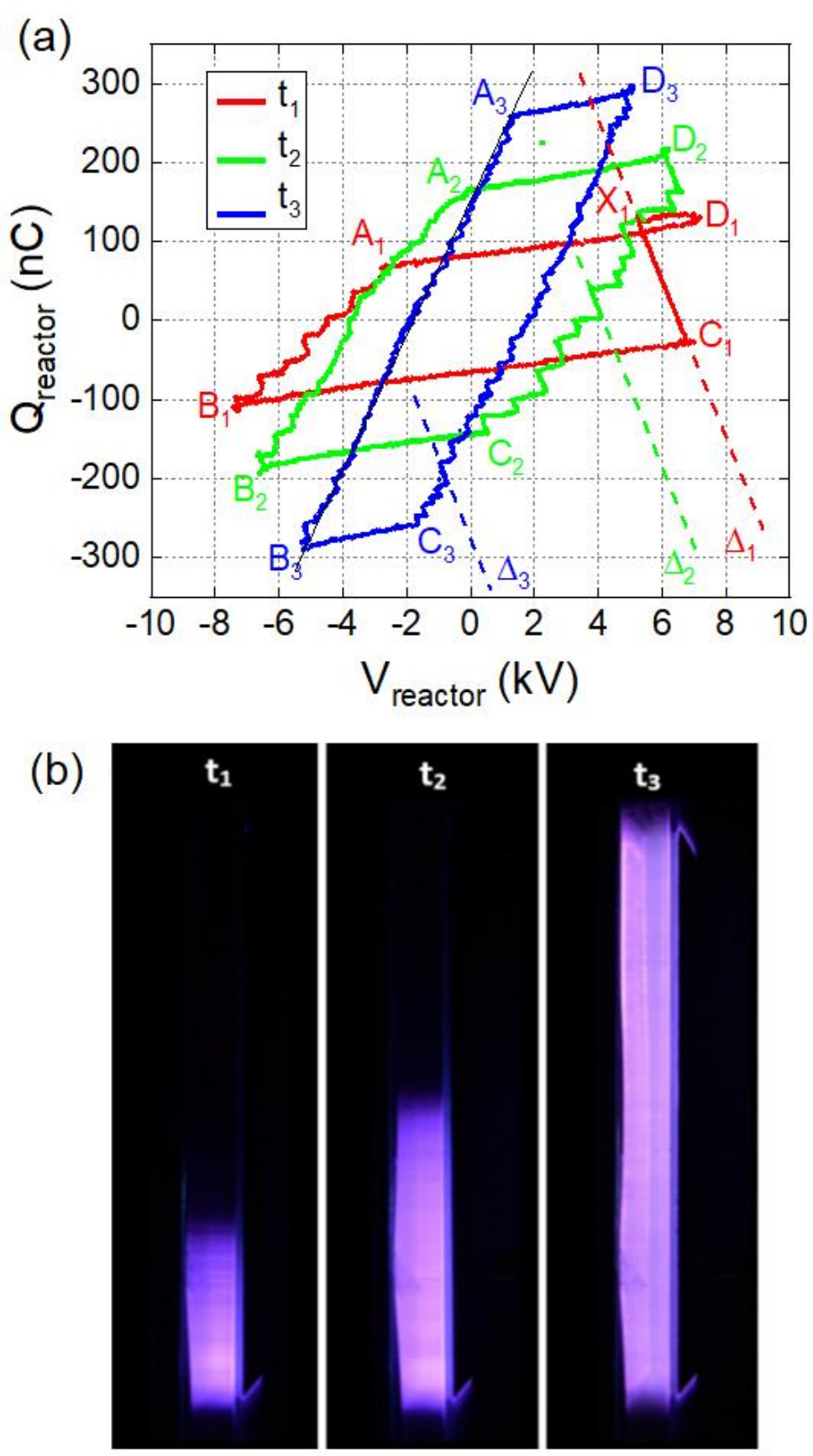

***Fig. 3 (a) (Q-V)$_{reactor}$ diagrams for three β values corresponding to different confinement volumes of plasma within the interelectrode region (b) Photographs of the plasma filling the gap region (12.0 cm height, 3.0 cm width) at different instants. Exp. conditions: 600 Hz, Helium 2 slm.***

We propose to experimentally verify the validity of the model through the relevance of the α and β parameters in the NB configuration. For this, the β values are measured in two different approaches:

- The first approach consists to plot the (Q-V)$_{reactor}$ diagram to deduce $\xi_{Off}$ and $\xi_{ON}$ and then to determine the values of α and β by solving the system $\begin{cases} \alpha + \beta = 1 \\ \xi_{ON} = \alpha.\xi_{OFF} + \beta.C_{diel} \end{cases}$. The Q$_{reactor}$ and V$_{reactor}$ parameters are measured at $t_1$ = 1s, $t_2$ = 2s, $t_3$ = 3s after injection of helium in the DBD. The corresponding (Q.V)$_{reactor}$ diagrams and discharges photographs are represented in Figure 3a and 3b. A helium plasma is gradually formed as the helium flow rate expels air from the reactor, resulting in same $\xi_{OFF}$ values but different $\xi_{ON}$ values. As reported in Table 2, the β values have been estimated to 35.3%, 57.6% and 90.5% at $t_1$, $t_2$ and $t_3$ respectively.
- The second approach consists to evaluate the parameters V' and V" defined in equation {2} and which represent the volumes without/with electrical charges in the PB-DBD. The volumes of the gap and of the plasma ($V_{gap}$ and V") are reported in Table 2, as well as the resulting β values: 31.7%, 50.0% and 95.8% at $t_1$, $t_2$ and $t_3$ respectively.

The β values obtained following approaches 1 and 2 show good correlation and sustain the viability of Peeter's model.

| | | $t_1$ = 1s | $t_2$ = 2s | $t_3$ = 3s |
|---|---|---|---|---|
| **Approach "1"** | $\xi_{OFF}$ $(pF)$ | 5.56 | 5.56 | 5.56 |
| | $\xi_{ON}$ $(pF)$ | 37.4 | 57.4 | 87.1 |
| | $V_{diel}$ $(cm^2)$ | 3.0×0.15×12.0 | 3.0×0.15×12.0 | 3.0×0.15×12.0 |
| | $C_{diel}$ $(pF)$ | 95.6 | 95.6 | 95.6 |
| | $\beta = \frac{\xi_{ON} - \xi_{OFF}}{C_{diel} - \xi_{OFF}}$ | 35.3% | 57.6% | 90.5% |
| **Approach "2"** | $V_{gap}$ $(cm^3)$ | 3.0×1.0×12.0 | 3.0×1.0×12.0 | 3.0×1.0×12.0 |
| | $V" = V_{plasma}$ $(cm^3)$ | 3.0×1.0×3.8 | 1.0×3.0×6.0 | 1.0×3.0×11.5 |
| | $\beta = \frac{V"}{V' + V"}$ | 31.7% | 50.0% | 95.8% |

***Table 2. Comparison of the β values obtained following approach "1" or approach "2" upon the slow formation of a He plasma within the gap volume.***

In Figure 3a., the three Lissajous diagrams do not have the same characteristics. Whatever the values of β (and therefore of time), the upper and lower sides remain always parallel ($A_1D_1$ // $B_1C_1$ // $A_2D_2$ // $B_2C_2$ // $A_3D_3$ // $B_3C_3$) with therefore $\xi_{OFF}$≈5 pF. If the parallelism of the lateral sides $B_xA_x$ and $C_xD_x$ is identical for each individual value of β, this is no more the case if these sides belong to different values of β (for example $A_1B_1$ is no more parallel to $A_3B_3$). This means that the $\xi_{ON}$ values depend on the β values, in accordance with equation {9} of Peeter's model.

The supplied voltage being sinusoidal, the micro-discharges are generated upon the activity phases of the negative/positive half-cycles, i.e. along the $A_xB_x$ and $C_xD_x$ sides. Q$_{reactor}$ is the same on $A_xB_x$ and $C_xD_x$ sides but is not necessarily "divided" by the same number of micro-discharges on each of these sides. At instant $t_1$, Q$_{reactor}$ is carried by 6 filaments on the negative half-cycle (or 6 segments with negative slope on the $A_1B_1$ side) but by a single filament on the positive half-cycle (or one segment with negative slope on $C_1D_1$ side). The electric charge of the filament along $C_1D_1$ is particularly high since its value is $|Q(C_1) - Q(X_1)| = |-30 - 130| = 160nC$. At $t_2$, the slopes of $A_2B_2$ and $C_2D_2$ increase and their respective numbers of micro-discharges increase as well. These trends are also confirmed at $t_3$, when plasma covers almost 95% of the interelectrode gap.

Whatever the value of $\beta^-$, the lines $\Delta_1$, $\Delta_2$ and $\Delta_3$ of the Lissajous diagram are parallel to each other, i.e. all the segments have the same negative slope. When its breakdown potential is reached, a micro-discharge propagates through the gap and can rebalance the charges on either side of the electrodes, as during the discharge of a capacitor. The sequential and cumulative nature of the charges which are deposited one after the other after crossing the gap has also been observed by Biganzoli et al. [45]. In our experimental conditions, the voltage is not constant during the very short lifetime of the micro-discharge: it decreases from the breakdown potential to a lower value, thus explaining why $\Delta V$ is negative and therefore why the segment presents a negative slope. This type of phenomenon has also been observed in various atmospheric pressure plasma jets (APPJ) and DBD [46] [47].

### III.A.2. Experimental validation of the $\gamma$ and $\delta$ parameters of the equivalent electrical model

$\gamma$ and $\delta$ represent the volume fraction of the gas and the volume fraction of the seeds respectively (see {4}). As these two parameters do not appear in the original model of Peeter, it is necessary to verify whether their integration in the present model is correct, especially through {7}. This relation stipulates that the total capacitance of the gap is equal to the sum of the capacitance of the gas (weighted by factor $\gamma$) and the capacitance of the seed bed (weighted by factor $\delta$).

For this, we carried out a first experiment consisting in gradually filling the interelectrode gap with corn seeds, going from 0 (no seeds) to 76 (fully filled gap) by steps of approx. 10 seeds. For each of these steps, we calculated the total volume occupied by the seeds, as indicated in Table 3. We also reported therein several electrical parameters using the LCRmeter: the $C_{gas}$ parameter (interelectrode gap containing only gas, i.e. without any seeds), the $C_{gap}$ parameter as well as the dissipation factor ($D_F$), the equivalent impedance (Z) and the phase angle ($\theta$). The parameter $\delta$ is deduced from the values of $V_{seeds}$ and $V_{gap}$ = 30cm³ in accordance with the relation {4}. Finally, the value of $C_{seeds}$ is deduced from the relation {7}.

Since the inter-electrode gap has a total volume of 30 cm³, its maximum capacity is limited to 76 seeds of corn. This configuration corresponds to a dielectric volume of 19.44 cm³ and a gas volume of 10.56 cm³. To obtain a volume of "corn material" close to the volume of the inter-electrode gap, i.e. to satisfy the condition $\delta$=1, 120 seeds were crushed and reduced into powder, and finally compacted into the interelectrode gap.

| $N_{seeds}$ (-) | $V_{seeds}$ (cm³) | $C_{gas}$ (pF) | $C_{gap}$ (pF) | $D_F$ (-) | Z (MΩ) | θ (°) | δ (-) | $C_{seeds}$ (pF) | Corn form |
|---|---|---|---|---|---|---|---|---|---|
| 0 | 0.00 | 2.80 | 2.80 | 0.004 | 25.4 | -89.7 | 0.000 | - | - |
| 11 | 2.77 | 2.80 | 4.39 | 0.024 | 24.2 | -88.6 | 0.093 | 19.89 | Seeds |
| 22 | 5.55 | 2.80 | 5.98 | 0.041 | 23.0 | -87.7 | 0.186 | 19.89 | Seeds |
| 33 | 8.33 | 2.80 | 7.57 | 0.053 | 21.7 | -86.9 | 0.279 | 19.89 | Seeds |
| 43 | 11.11 | 2.80 | 9.15 | 0.063 | 20.4 | -86.3 | 0.371 | 19.91 | Seeds |
| 54 | 13.88 | 2.80 | 10.74 | 0.071 | 19.3 | -85.9 | 0.464 | 19.91 | Seeds |
| 65 | 16.66 | 2.80 | 12.33 | 0.075 | 18.1 | -85.7 | 0.557 | 19.91 | Seeds |
| 76 | 19.44 | 2.80 | 13.92 | 0.078 | 16.8 | -85.6 | 0.650 | 19.90 | Seeds |
| 120 | 30.00 | 2.80 | 20.51 | 0.082 | 14.5 | -85.2 | 1.000 | 20.95 | Powder |

***Table 3. Electrical parameters of the PB-DBD measured with LCRmeter. The device is filled by corn either in seeds form or in powder form.***

From the values in Table 3, it is possible to plot $C_{gap} - \gamma_{Corn}.C_{gaz}$ as a function of the $\delta$ parameter, as shown in Figure 4a. The curve obtained is a straight line whose slope corresponds to $C_{seeds}$ for the different values of $\delta$. The value of $C_{seeds}$ can also be read by projecting on the Y-axis the point at coordinate $\delta$ = 1. The mean value of $C_{seeds}$ is estimated to 19.32 pF ($\pm$ 0.08 pF). In overall, the quality of the linear fit is satisfying since its coefficient of determination $\rho^2$ has a value higher than 0.99. When corn is reduced into powder, the value of $C_{seeds}$ is slightly higher than what was expected (20.98 pF). Two reasons may explain this discrepancy: (i) the relative humidity of the ambient atmosphere has stronger influence when corn is under powder due to higher surface-to-volume ratio, (ii) the grinding and compacting processes may not form a powder with grains sufficiently small in size and might also leave metallic traces. The figure 4b presents the $C_{seeds}$ values of the 6 types of seeds studied in this article. A first group of seeds has a value close to 14 pF (radish, coriander, sunflower) while the other has a value close to 20 pF (lentil, bean, corn).

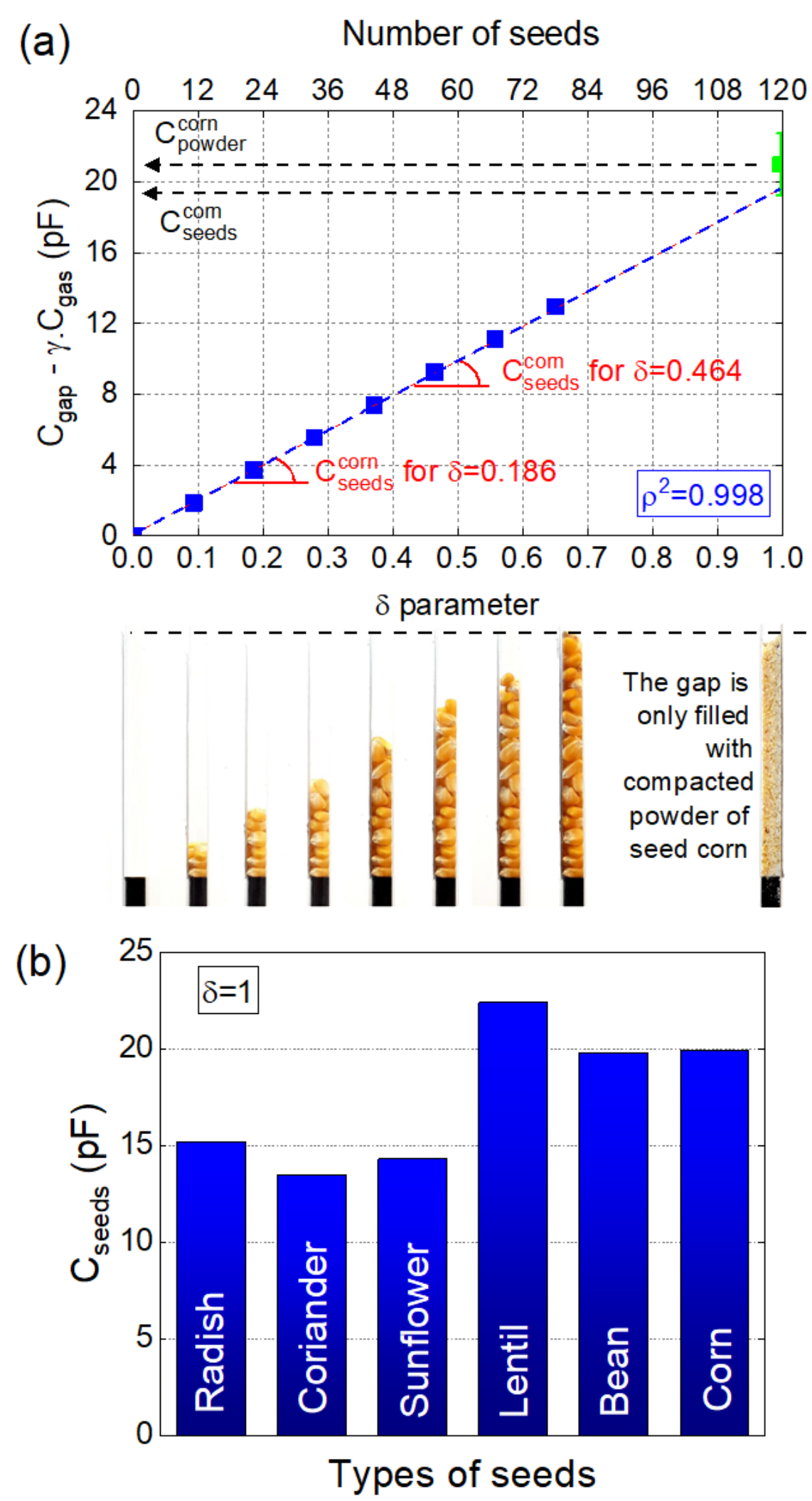

***Figure 4. (a) Variation of $C_{gap} - \gamma_{Corn}.C_{gaz}$ as a function of the $\delta$ parameter for different beds of corn seeds filling the DBD, (b) Estimation of seeds bed capacitances for six different types of seeds for $\delta$=1. All these measurements are carried using an LCRmeter while plasma is off.***

According to the Table 3, a higher number of seeds in the packed-bed DBD induces a lower impedance, going from 25.4 to 16.8 MΩ as well as a decrease in phase angle from -89.7° to -85.6°. In our equivalent electrical model, the seed bed is represented by a single capacitor and not by the parallelization of a capacitor and a resistor. This approximation is correct as long as the equivalent resistor has a very high (or quasi-infinite) value. Here, this situation is considered acquired since all the seeds had equivalent series resistances values greater than 10 MΩ. However, a seed bed will not systematically have an almost infinite equivalent series resistance, especially if the seeds have a very high-water content. In that case, a capacitor in parallel with a resistor will be a mandatory for realistic modeling. Finally, if the plasma treatment times are prolongated, one can expect a gradual loss of seeds water content and therefore a variation of capacitance (and resistance) of the seed bed. Here again, correct modeling will rely on the paralleling of a capacitor and a resistor.

### III.A.3. Influence of oxygen flow rate on the discharge electrical parameters

To understand the interdependence of the electrical and chemical properties of plasma, the NB-DBD has been supplied with a mixture of helium (2 slm) and molecular oxygen (0-100 sccm). The influence of increasing $O_2$ flow rates ($\Phi_{O2}$) on the discharge electrical properties has been evaluated through the following electrical parameters: $\xi_{OFF}$ (capacitance of the DBD during the quenching phase) and $\xi_{ON}$ (capacitance of the DBD during the activity phase) in Figure 5a, $V_{br}$ (breakdown voltage) in Figure 5b, β (fraction of domains containing electrical charges) in Figure 6a, $P_{plasma}$ (electrical plasma power) in Figure 7a and $Q_f$ (electrical charge per filament) in Figure 7b. All these electrical parameters can change for $O_2$ flow rates higher than 5 sccm except $Q_f^-$ which remains constant with a value of 12 nC whatever $\Phi_{O2}$. The plasma power remains close to 1.2 W between 0 and 5 sccm, then increases up to 1.6W at 15 sccm before decreasing to 0.7W at 100sccm.

Certain electrical parameters have diverging values, depending on they are measured during a positive or negative half-cycle:

- From 0 to 50 sccm, the Figure 5a indicates that $\xi_{ON}^-$ equals 88 pF: a value close to the barrier capacitance: $C_{diel}$ = 90pF. Then, $\xi_{ON}^-$ decreases to 81 pF (at 70 sccm) and to 60 pF (at 100 sccm) since an increase in the $O_2$ flow rate reduces the distribution of electrical charges into the gas volume. This is particularly visible in Figure 6b, for $\Phi_{O2}$ increasing from 50 to 100 sccm. When the number of micro-discharges occurring on the positive (or anodic) half-cycle becomes too low, the extraction of the corresponding $\xi_{ON}^+$ value (from the Lissajous curve) becomes less accurate.
- The breakdown voltage increases with $\Phi_{O2}$ since it changes from 2.0 kV (0 sccm) to 5.9 kV (100 sccm) on the negative half-cycle and from 2.0 kV (0 sccm) to 7.1 kV (100 sccm) on the positive half-cycle.
- The $\beta^-$ and $\beta^+$ parameters follow the same trend as $\xi_{ON}^-$ and $\xi_{ON}^+$ to satisfy formula {5}. The Figure 6b shows some photographs of the discharge operating at different oxygen flow rates. At 0 and 20 sccm, plasma appears more emissive than in the other pictures, fills the entire gap volume and therefore fully covers the electrodes surfaces. At 70 and 100 sccm, plasma does not cover the entire electrodes anymore. The gap volume has both dark and emissive regions, the latter ones being characterized by micro-discharges along the interelectrode gap.
- In Figure 7b, while the electrical charge per filament $Q_f^-$ remains fixed at 12 nC during the negative half cycle whatever the oxygen flow rate, $Q_f^+$ follows another trend: it increases from 12 nC to 50 nC on the 2-50 sccm range and decreases from 50 nC to 35 nC after. Comparing the positive and negative components of the $Q_f$ parameter leads to the conclusion that upon positive half-cycle, micro-discharges are lower in number although more intense in magnitude. Such dependence on half-cycle polarization has already been observed in the literature [48], [49], [50]. The increasing divergence between $Q_f^-$ and $Q_f^+$ is discussed in section III.B.2.
-

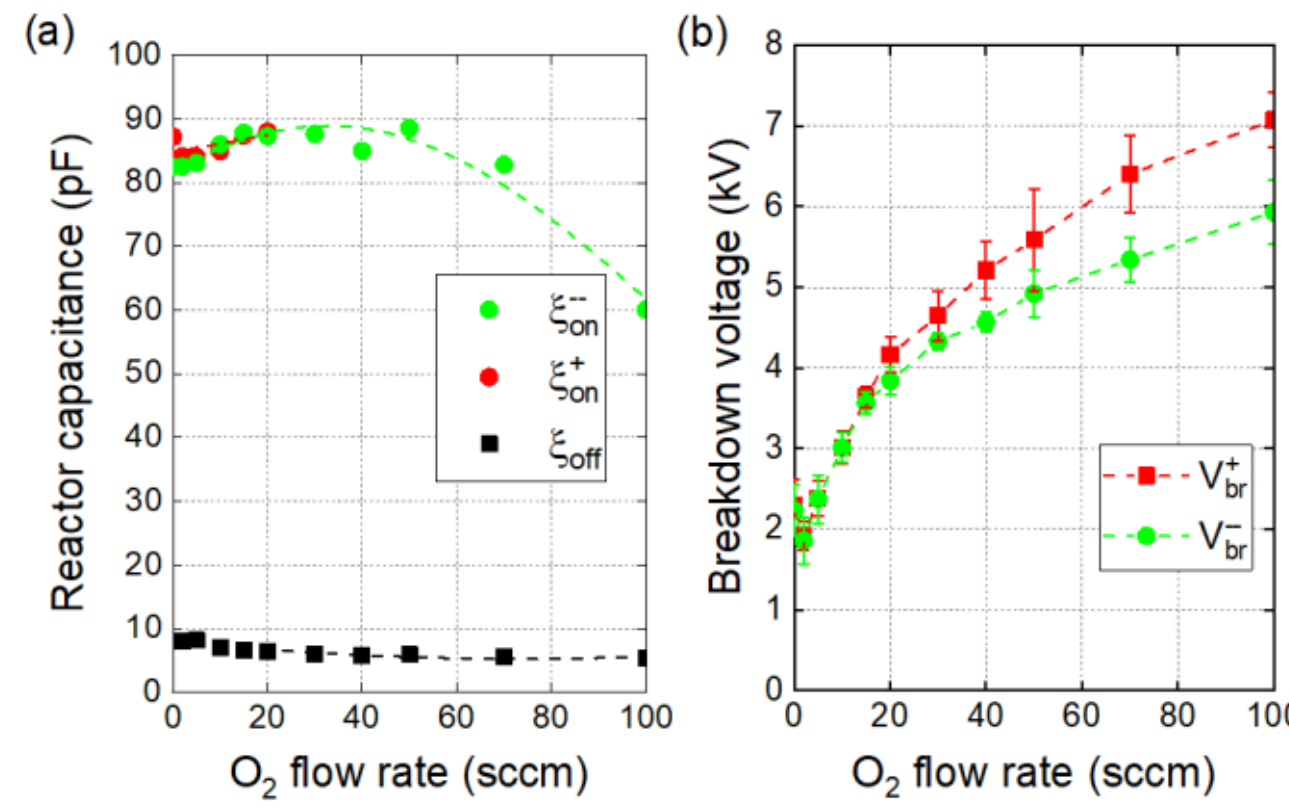

***Fig. 5 (a) Electrical equivalent capacitance of the reactor vs oxygen flow rate, without discharge ($\xi_{OFF}$) or with discharge ($\xi_{ON}$) (b) Breakdown voltage ($V_{br}$) vs oxygen flow rate. In both figures, the superscript "-" and "+" signs stand for negative and positive half cycles respectively.***

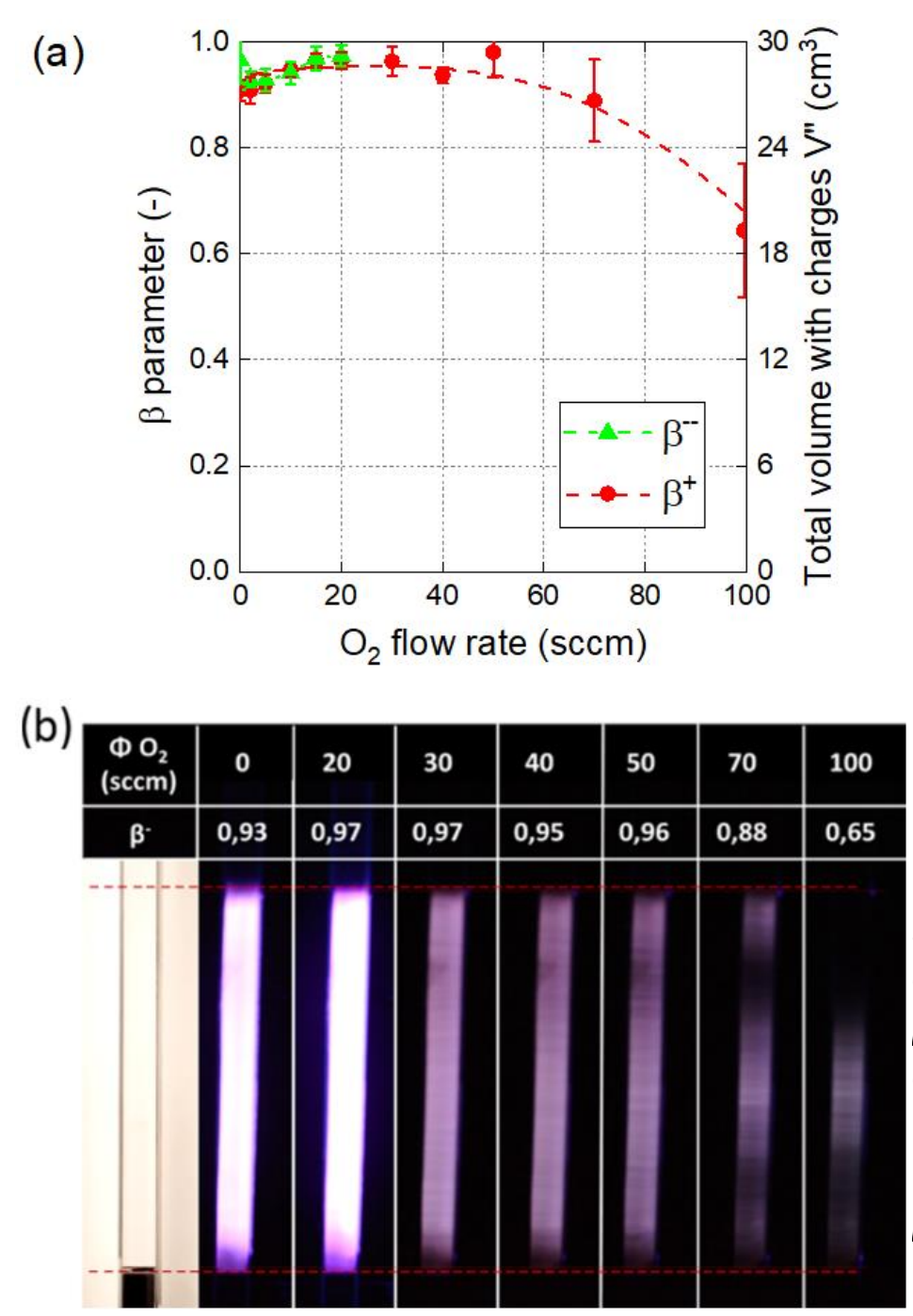

***Fig. 6 (a) Positive and negative β parameters assessed for a He-$O_2$ discharge and (b) related pictures.***

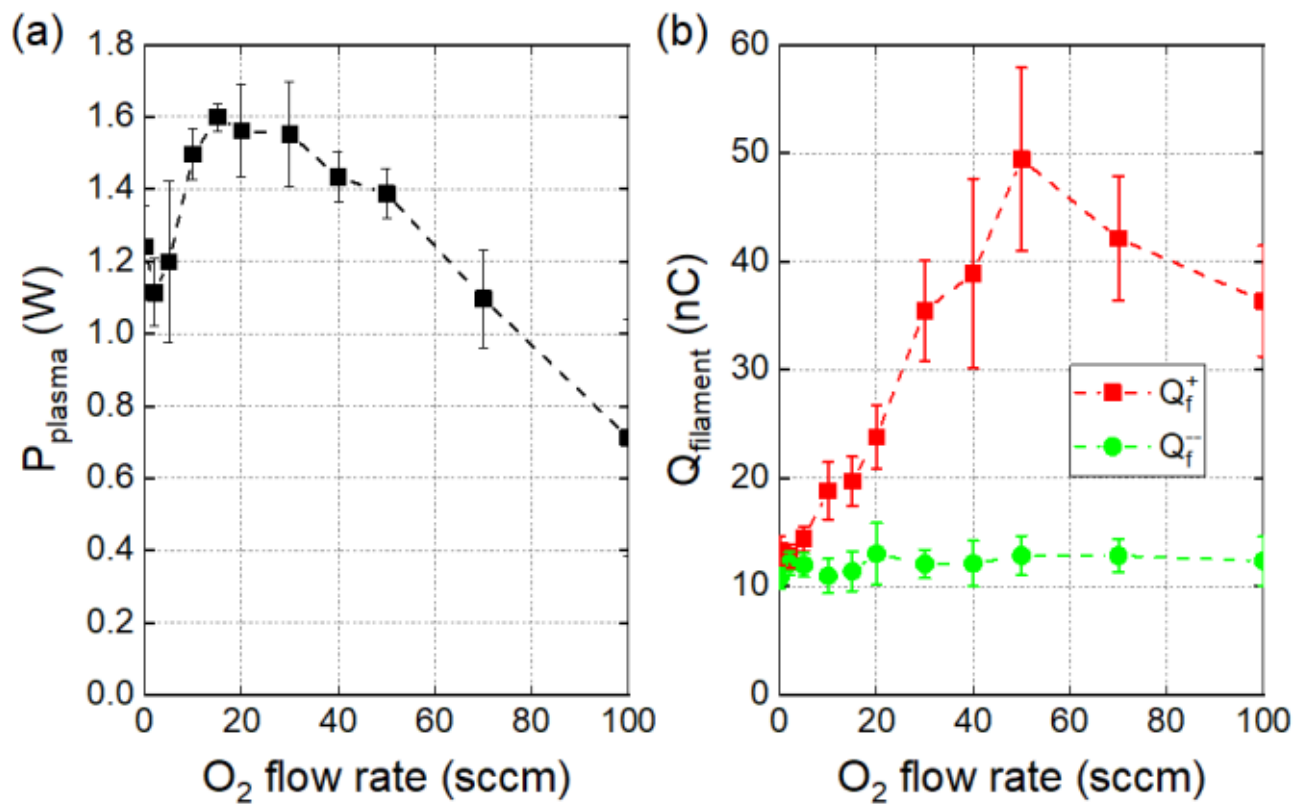

***Fig. 7 Influence of oxygen flow rate on (a) electrical plasma power and (b) electrical charge per filament during positive (anodic) half cycle and negative (cathodic) half cycle.***

## III.B. Plasma electrical parameters in the PB-DBD reactor

### III.B.1. Influence of seeds specimens and $O_2$ flow rate on plasma electrical parameters

The influence of six seeds specimens (beans, radishes, corianders, lentils, sunflowers and corns) has been investigated on the electrical parameters of the discharge operating in the packed bed configuration supplied in helium (2 slm) with/without molecular oxygen (0-100 sccm). Figure 8a shows the $\xi_{OFF}$ capacitance as a function of $\Phi_{O2}$ for the 6 types of seeds and for the NB configuration. In all cases, the decreased in $\xi_{OFF}$ is observed if $\Phi_{O2}$>5sccm. Then, the curves corresponding to the 6 packed-bed configurations can be dispatched in two groups: the group I includes radishes, sunflowers and corianders while the group II is composed of lentils, beans and corns. Pack-bedding the DBD reactor with seeds make its $\xi_{OFF}$ capacitance always higher than in the NB configuration (5 pF). Hence, for $\Phi_{O2}$ increasing from 0 to 100 sccm, $\xi_{OFF}$ decreases from 10.1 pF to 8.8 pF for group I while it decreases from 14.2 pF to 11.7 pF for group II. Moreover, the Figure 8a clearly shows that the $\xi_{OFF}$ variations are more important for group II than for group I.

The distribution of seeds into 2 groups is also relevant if we consider other electrical parameters such as breakdown voltages. As shown in Figure 8b, this separation is even more marked as $\Phi_{O2}$ increases until reaching the value of 100 sccm. Then the mean value of $V_{br}$ is close to 6.8 kV and 5.0 kV for group I and group II respectively. The half-cycle polarities induce the same values of $V_{br}^+$ and $V_{br}^-$ which, therefore, are not represented in the figure.

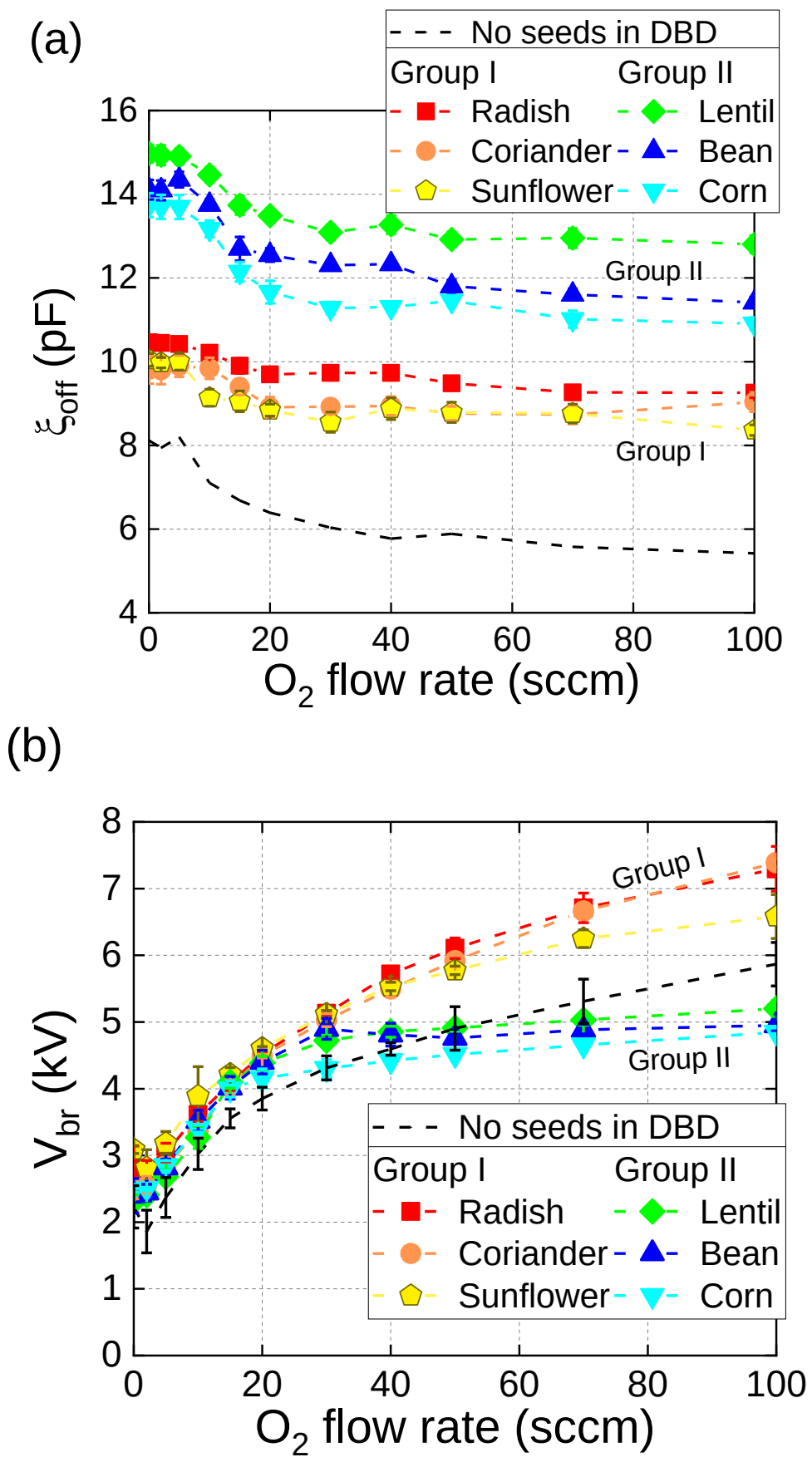

***Fig. 8 Influence of seeds on (a) equivalent capacitance without plasma $\xi_{OFF}$ and (b) breakdown voltage.***

In Figure 9a, the calculation of $\beta^-$ with equations {3} and {9} leads to a value as high as 0.95 for $O_2$ flow rate in the range 0-50 sccm for group I and 0-15 sccm for group II. Beyond, $\beta^-$ decreases to values slightly higher than 0.3 for group I and lower than 0.3 for group II. As shown in Figure 9a, $\beta^-$=0.3 corresponds to V"≈3 cm³ where V" stands for the total volume occupied by the charges within the gap electrode and therefore to the total volume of contact points regions. To illustrate the relevance of the computed $\beta^-$ parameter, the Figure 9b shows photographs of the DBD packed with corn seeds and supplied with $\Phi_{O2}$ = 0-100 sccm.

In PB-DBD supplied with helium at 2 slm, the optical emission of plasma is quite uniform in the entire gap volume, meaning that the micro-discharges follow a random distribution throughout the interelectrode gap. The admixture of molecular oxygen induces a decay in the overall discharge emissivity as well as a more heterogeneous discharge. Gaseous interstices, resulting from the stacking of the seeds in the gap volume, can be either non-ionized (local regions of low electric field) or ionized (local regions where electric field is high enough to generate a cold plasma). Several determining criteria can explain why plasma is preferentially formed in some interstices: (i) interstice critical size, (ii) local peak effects due to seeds topography and (iii) existence of triple junctions like "seed - dielectric barrier - plasma gas" or "seed - electrode - plasma gas". According to Figure 9b, these two types of triple junctions are equally plausible since for increasing values of $\Phi_{O2}$, the discharge only appears on the seeds outer contour in contact with the dielectric barrier and the inner plate electrode of the DBD. In such configuration, the discharge can be considered as surfacic or polar.

Overall, Figure 10a shows that plasma electrical power is the same with/without packed-bed seeds with values close to 1.2 W for $\Phi_{O2}$<10sccm and as low as 0.6W at 100sccm. In PB configuration at $\Phi_{O2}$<50 sccm, the seeds from group I induce an electrical plasma power 63% higher than seeds from group II. In Figure 10b, the (Q-V) diagrams have been plotted for $\Phi_{O2}$ = 50 sccm. The lateral sides of the Lissajous figures have slopes (or $\xi_{ON}$ values) which strongly depend on the types of seeds. Again, the two seed groups can be formed with $\xi_{ON}$ = 64.3 pF for group I and $\xi_{ON}$ = 36.0 for group II.

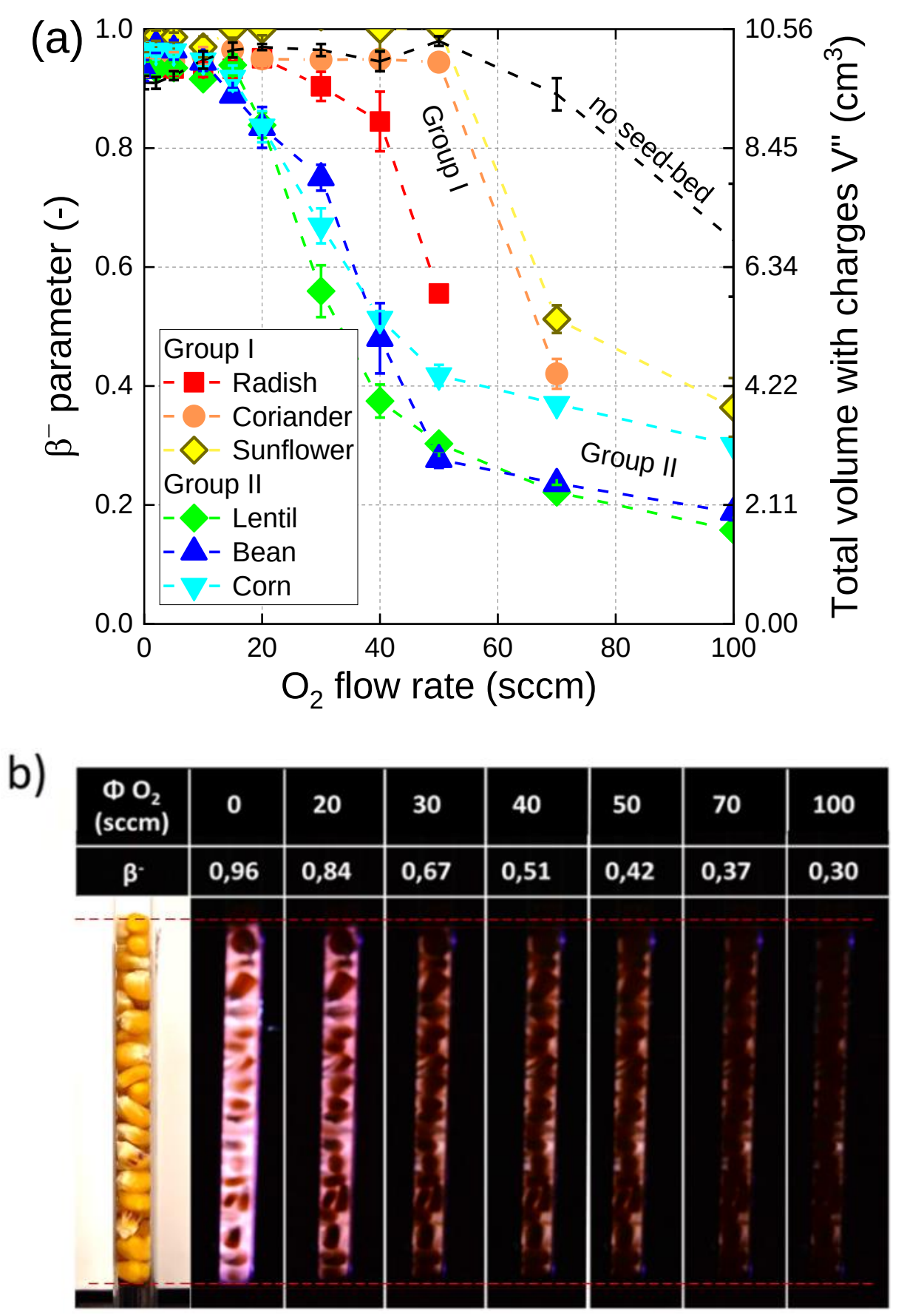

***Fig. 9 (a) Electrode area fraction covered with plasma (β) vs O₂ flow rate for 6 specimens of seeds (b) Pictures of the PB-DBD reactor filled with corn seeds and observed for different O₂ flow rates.***

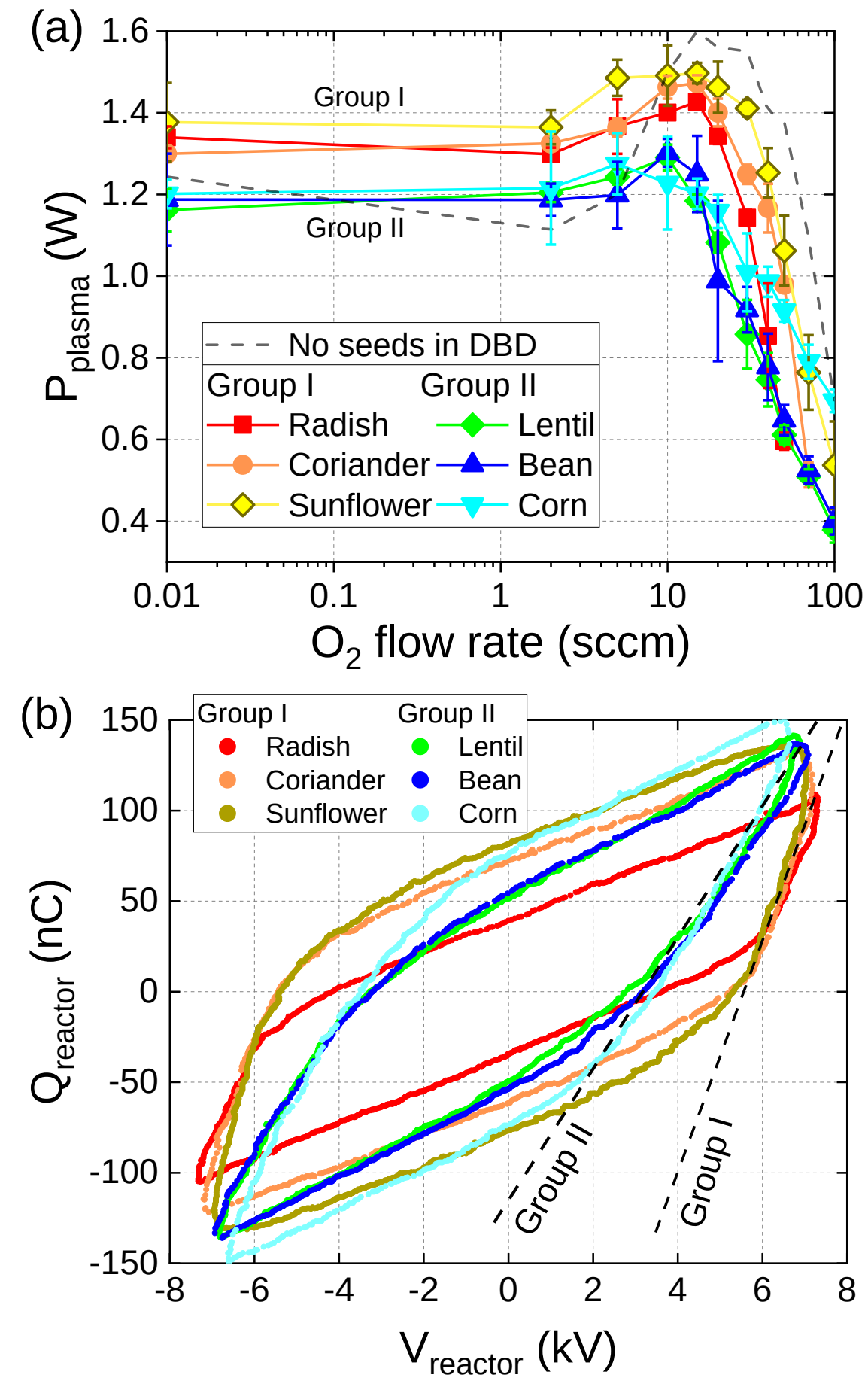

***Fig. 10 (a) Plasma electrical power as a function of O₂ for different seeds specimens, (b) (Q-V)$_{reactor}$ diagrams for different seeds specimens with the packed bed DBD reactor supplied with 2slm and 50 sccm of helium and oxygen respectively.***

Finally, optical emission spectroscopy has been achieved to characterize the main emissive species generated in the seed-packed DBD reactor supplied with a helium flow rate of 2 slm and $O_2$ flow rates increasing from 0 to 100 sccm. In Figure 11, these measurements indicate that the admixture of molecular oxygen does not drive to the production of radiative OH or H radicals. Unsurprisingly, only radiative O radicals can be detected (777.2 nm) as well as molecular nitrogen from the second positive system ($N_2^*$ at 337.1 nm and $N_2^+$ at 391.4 nm). Overall, while the emission of helium and nitrogen species decay as a function of $\Phi(O_2)$, the emission of O radicals shows a local maximum close to 10 sccm. This maximum may be related with the maximum of electrical plasma power evidenced in Figure 10. For $\Phi(O_2)$>10 sccm, all the emissions clearly decay, potentially leading to the production of non-radiative species like ozone.

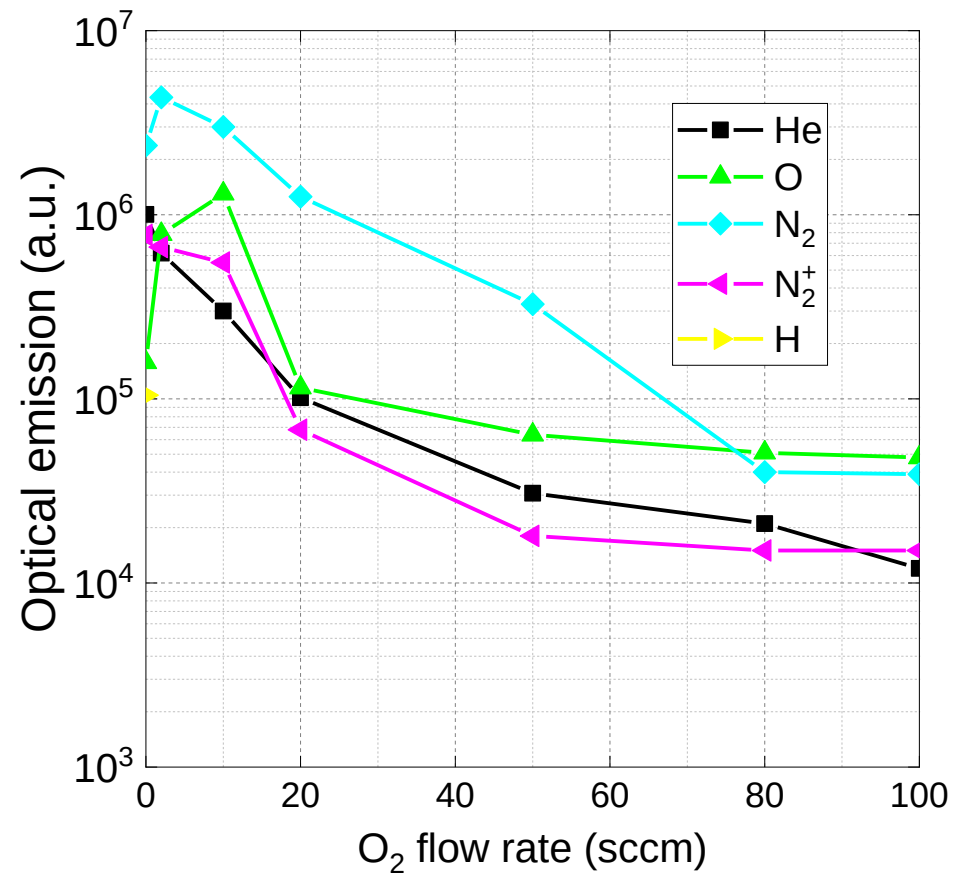

***Fig. 11. Influence of $O_2$ flow rate on the optical emission of species detected in the seedpacked DBD reactor (helium flow rate at 2 slm).***

### III.B.2. How seeds and molecular oxygen mitigate the micro-discharges distribution

Figure 12 shows the time profiles of the discharge voltage and current for the DBD reactor in NB and PB configurations, supplied with helium (2 slm) with/without oxygen (50 sccm). Over a single period, the voltage shows an anodic (positive) half-cycle and a cathodic (negative) half-cycle. Each of these half-cycles is composed of activity and quenching phases as depicted in Figure 12. The discharge current appears as a distribution of peaks occurring upon the activity phases.

In the *He/NB configuration* and during the activity phase of the cathodic half-cycle, the micro-discharges are initiated from the dielectric barrier and appear as a set of peaks of low intensity, hence explaining the diffuse character of plasma. Then, the activity phase of the anodic half-cycle is characterized by a distribution of micro-discharges (i) initiated from the inner electrode, (ii) fewer in number and (iii) with much higher intensities to maintain an elevated transfer of the electrical charges.

From *He/NB to HeO₂/NB configuration*, two remarks are noteworthy:

- The discharge mechanisms occurring upon each activity phase depend on the gas nature. The admixture of molecular oxygen (electronegative gas) to helium increases the breakdown potential [51] and, consequently, drives to a shorter activity phase. While this phase has a typical duration of 0.50 ms in helium (without $O_2$ added), it decreases down to 0.35 ms in a He-$O_2$ plasma. Since this later ignition of the micro-discharges induces a shorter time window, these micro-discharges are fewer in number but show higher magnitude to maintain the total charge constant, as shown in Figure 12.
- In the presence of oxygen, $V_{br}^+ > V_{br}^-$. For this reason and as illustrated in Figure 7b, the charge per filament $Q_f^+$ increases dramatically with oxygen during anodic half-cycle while $Q_f^-$ remains unimpacted upon cathodic half-cycle. For high $O_2$ flow rates, the plasma discharge can cover an area smaller than the electrode surface, as indicated by the $\beta^-$ values in Figure 6b.

*From NB to PB configuration*, i.e. if the DBD reactor is pack-bedded with seeds, plasma discharge mechanisms are changed upon the active phase of the anodic half-cycle: as shown in Figure 12, the number of peaks is increased while their overall intensities are reduced. We assume that in comparison with the NB configuration, the seeds layout in the gap volume reduces the magnitude of electric field in the gaseous interstices hence the low-magnitude current peaks. At the same time, the presence of seeds induce a higher surface-to-volume ratio responsible for the higher number of micro-discharges, resulting in a surfacic (or polar) plasma, as depicted in Figure 9b and as already observed in the literature [39]. Finally, one must stress that no correlation was found between seeds' shape or size to explain the electrical differences between the two groups. Such distinction may be explained by the seeds intrinsic capacitances, as detailed in section III.D.

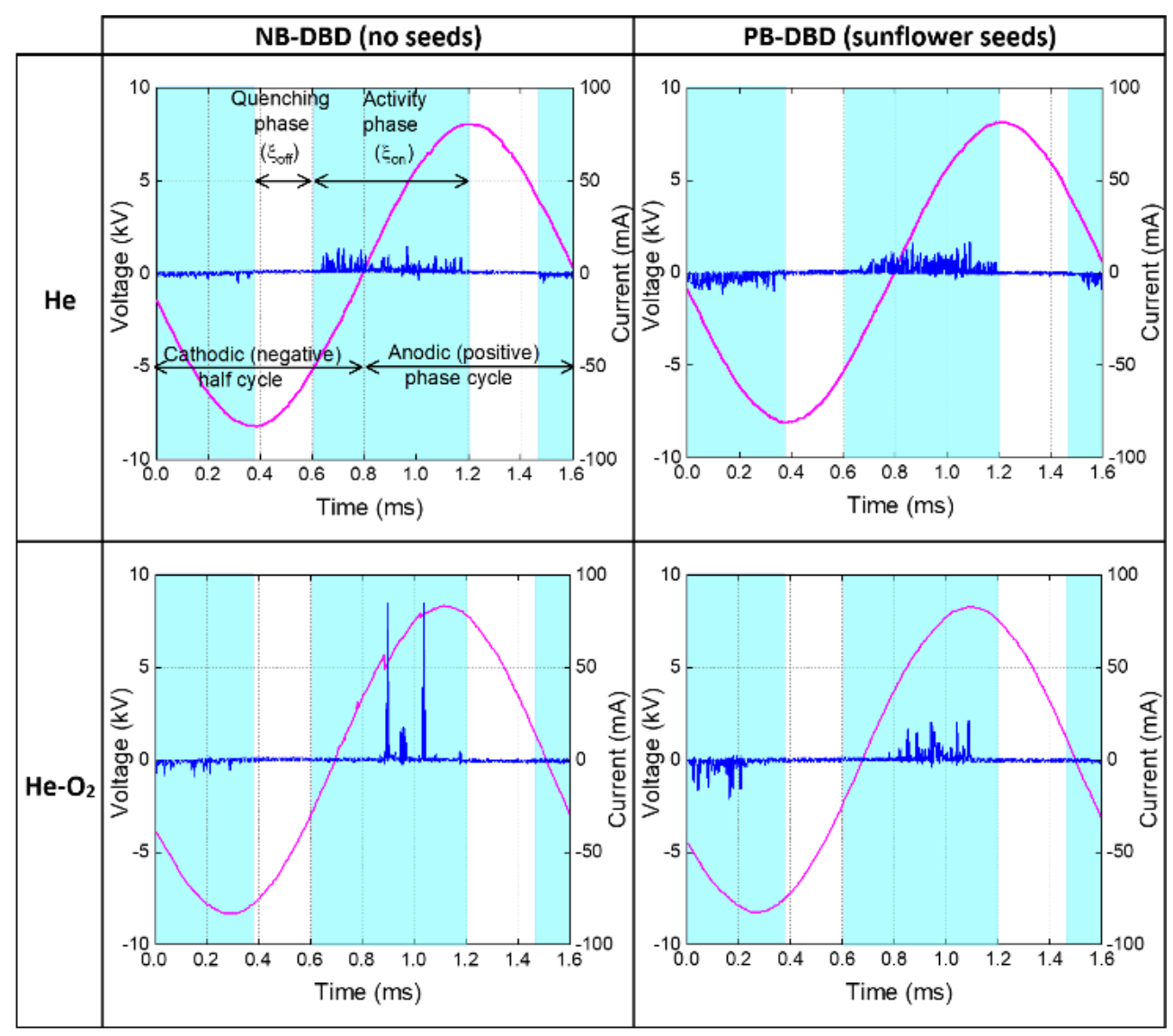

***Fig. 12 Time profiles of HV generator voltage and plasma current in NB-DBD reactor supplied with He at 2 slm and $O_2$ at 50 sccm.***

## III.C. Assessing the effects of plasma on seeds germination

Several types of seeds (lentils, radishes, sunflowers, beans, corianders and corns) have been treated by the PB-DBD process to investigate the benefits of plasma on seeds germinative properties. All the seeds have been stored in ambient air and in a dark enclosure where temperature and relative humidity were 22°C (±1°C) and 40% (±5%) respectively. These seeds have been observed with a stereo microscope (Optika, SZM-2) as shown in Figure 13. Their weights were assessed with an analytic scale (Sartorius, Entris 124i-1S) and their dimensions (length, width and thickness) estimated with caliper as reported in Table 4. For each type of seed, the characteristic dimension (L*) is calculated as the average of length ($L_S$), width ($W_S$) and thickness ($T_S$) while the aspect ratio is defined as the ratio between the highest to the lowest value among length, width and thickness. The volume of a single seed is determined by relation {15} while $V_{seeds}$ (product of $V_{1seed}$ by the number of seeds packed in the DBD) is given by {16}. Finally, the $\gamma$ and $\delta$ parameters can be calculated using the relations {4} and {5}.

$$V_{1seed} = \frac{4}{3}\pi\left(\frac{L_s}{2} \times \frac{W_s}{2} \times \frac{T_s}{2}\right)^3 \quad \{15\}$$

$$V_{seeds} = V_{1seed} \times N_{seeds} \{16\}$$

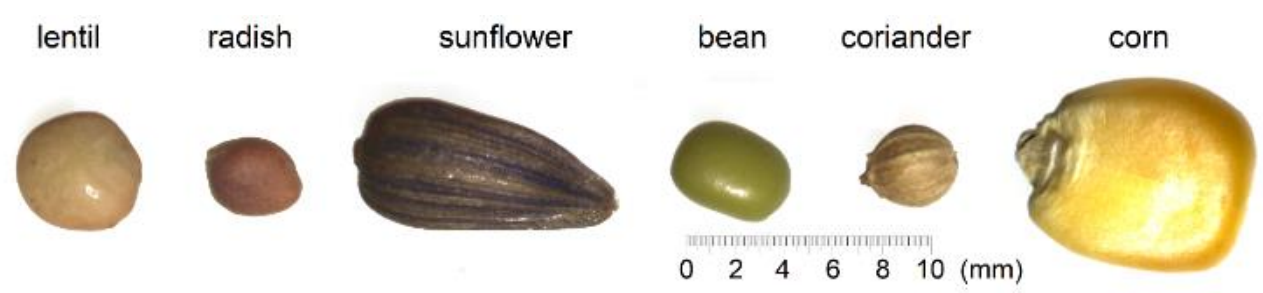

***Fig. 13. Seeds of lentil, radish, sunflower, bean, coriander and corn observed with the stereo microscope.***

| | GROUP I | | | GROUP II | | |
|---|---|---|---|---|---|---|
| | Radish | Cori -ander | Sun -flower | Lentil | Bean | Corn |
| Weight (mg) | 14.31 | 10.07 | 59.84 | 30.71 | 61.99 | 336.8 |
| Length (mm) | 3.84 | 3.14 | 11.33 | 4.55 | 4.08 | 10.40 |
| Width (mm) | 2.97 | 3.14 | 5.17 | 4.43 | 3.95 | 7.94 |
| Thickness (mm) | 1.98 | 3.14 | 3.26 | 2.26 | 5.35 | 5.92 |
| L* (mm) | 2.93 | 3.14 | 6.59 | 3.75 | 4.46 | 8.09 |
| Aspect ratio (-) | 1.94 | 1.00 | 3.48 | 2.01 | 1.35 | 1.76 |
| $V_{1seed}$ ($mm^3$) | 11.82 | 16.20 | 99.93 | 23.84 | 45.12 | 255.83 |
| $N_{seeds}$ | 1588 | 1192 | 216 | 848 | 452 | 76 |
| $V_{seeds}$ ($cm^3$) | 18.77 | 19.31 | 21.59 | 20.22 | 20.40 | 19.44 |
| $\gamma$ | 0.37 | 0.36 | 0.28 | 0.33 | 0.32 | 0.35 |
| $\delta$ | 0.63 | 0.64 | 0.72 | 0.67 | 0.68 | 0.65 |

***Table 4. Weight and dimensional parameters of seeds exposed to plasma.***

For each experimental condition and whatever the type of seeds studied: (i) 5 tests were carried out by considering samples of 100 seeds each time, (ii) after their imbibition with tap water, the seeds were immediately sown. Figure 14a shows the time tracking of germinating bean seeds considering two cases: a "control" group where seeds have not been exposed to plasma and a "plasma" group where seeds were treated in PB-DBD for 2 minutes in He-O2 plasma (2slm-40sccm ). The average time to obtain 50% of the germinated seeds on an entire crop is estimated using the $T_{50}$ parameter, as shown in Figure 14a. The comparison between $T_{50}^C$ and $T_{50}^P$ (for control and plasma groups respectively) evidence the effect of plasma since a gain of time (vigor) of approximately 3 hours is obtained; a value which is significant compared to the values of $T_{50}^C$ and $T_{50}^P$. To better evaluate this gain, one can express the relative variation of the $T_{50}$ parameter using the equation {17} and thus verify whether the vigor is shortened ($\Delta T_{50} < 0$) or on the contrary prolonged ($\Delta T_{50} > 0$):

$$\Delta T_{50} = \frac{T_{50}^P - T_{50}^C}{T_{50}^C} \quad \{17\}$$

The Figure 14b reports this parameter for the 6 seeds specimens. It turns out that plasma treatment is seed-dependent since the seeds exhibit different biological responses to the same plasma treatment. While the dry plasma process reveals quite effective for lentils and beans seeds (with p-values lower than 0.05), it seems less relevant for sunflower and corn seeds and useless for radish and coriander since their $\Delta T_{50}$ values are statistically too close to the control reference ($\Delta T_{50}$ = 0) to consider plasma triggered biological effects. Overall, we can only assert that seeds with the highest capacitances (approx. 20 pF) recorded the largest variations of $\Delta T_{50}$ (on the order of -11%). Besides, no correlation has been found between $\Delta T_{50}$ and seeds dimensions (either aspect ratio or characteristic dimension) as reported in Figure 15 or even between $\Delta T_{50}$ and seeds weights. For these reasons, it seems reasonable to hypothesize that the more a seeds' bed will have a high capacitance, the more the plasma process will release dormancy, probably thanks to higher values of electric field at seeds' contact points.

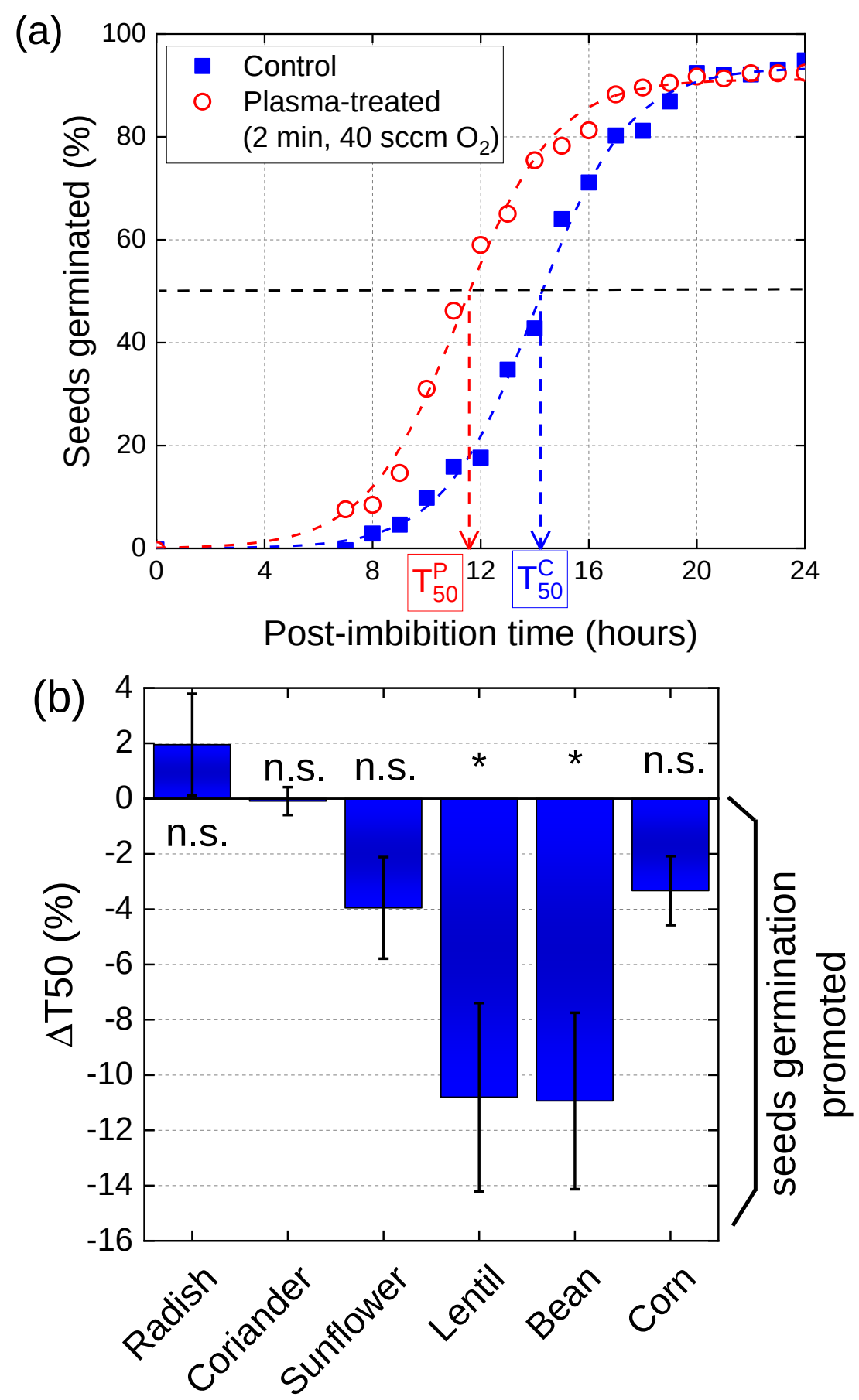

***Fig. 14 (a) Germination curves of untreated and plasma-treated seeds of beans (b) Relative variations of the $T_{50}$ parameter for the 6 types of seeds exposed to plasma, compared with control (b). Experimental conditions: plasma exposure time = 2 min, helium flow rate = 2 slm, $O_2$ flow rate = 40 sccm, data expressed as mean ± sem. P-value *<0.05.***

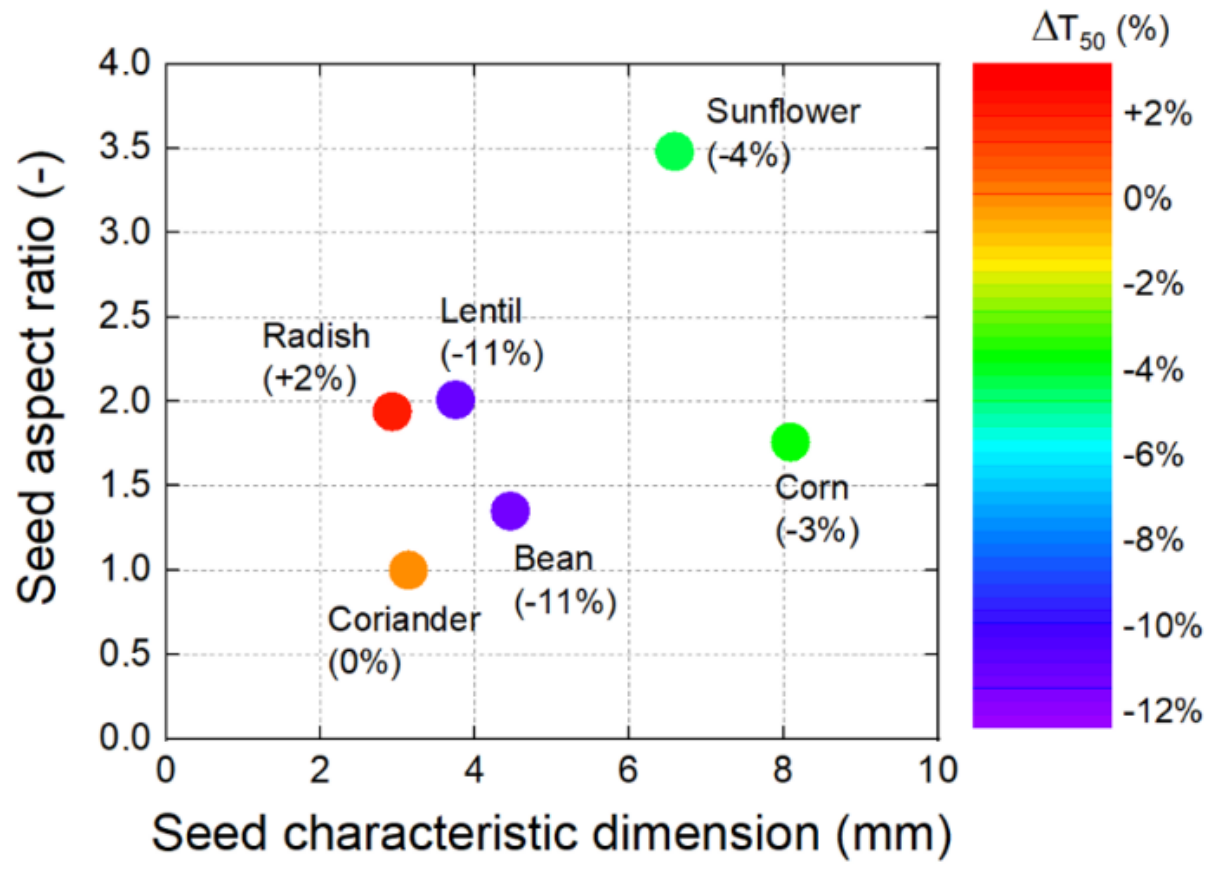

***Fig. 15. $\Delta T_{50}$ parameters of seeds reported in seed dimensional diagram (aspect ratio vs characteristic dimension).***

### III.D. Assessing the overall capacitance of packed-bed seeds

The overall capacitance of packed-bed seeds ($C_{seeds}$) can be evaluated through the following procedure:

(i) For each type of seeds packed in a DBD reactor, its $\xi_{off}$ value can be evaluated by measuring the slopes of the upper/lower sides from the (Q-V) diagrams in Figure 10b.
(ii) Since $C_{diel}$ = 90 pF (relation {6}), the corresponding values of $C_{gap}$ can be assessed with relation {8}.
(iii) Considering that gas is contained in the gap region (area = 3cm×10cm, gap = 10mm) of the PB-DBD, its capacitance is estimated to $C_{gas} \approx 2.65$ pF
(iv) The relation {7} provides the values of $C_{seeds}$ knowing the $\xi_{off}$, $C_{gap}$ and $C_{gas}$ values.

The equivalent electric model proposed in section II.B. enables to determine the capacitance of the packed-bed seeds, as reported in Table 5. Unsurprisingly, the seeds of group 1 have a capacitance with an average value of 14.2±0.1 pF while those of group 2 have a capacitance of 20.4±1.4 pF.

| | GROUP I | | | GROUP II | | |
|---|---|---|---|---|---|---|
| | **Radish** | **Cori -ander** | **Sun -flower** | **Lentil** | **Bean** | **Corn** |
| $\gamma$ | 0.37 | 0.36 | 0.28 | 0.33 | 0.32 | 0.35 |
| $\delta$ | 0.63 | 0.64 | 0.72 | 0.67 | 0.68 | 0.65 |
| $\xi_{off}$ (pF) | 14.31 | 10.07 | 59.84 | 30.71 | 61.99 | 336.8 |
| $C_{gap}\ (pF)$ | 9.99 | 9.96 | 10.88 | 15.58 | 14.39 | 13.44 |
| $C_{seeds}^{model}\ (pF)$ | 14.3 | 14.1 | 14.1 | 21.9 | 19.9 | 19.3 |
| $C_{seeds}^{LCRmeter}\ (pF)$ | 15.2 | 13.5 | 14.3 | 22.4 | 19.8 | 19.9 |
| $\left\lvert\frac{C_{seeds}^{LCRmeter} - C_{seeds}^{model}}{C_{seeds}^{LCRmeter}}\right\rvert$ | 5.92% | 4.44% | 1.40% | 2.23% | 0.51% | 3.02% |

***Table 5. Capacitance of the PB-DBD when plasma is off, of the gap and of the seeds.***

## IV. Conclusion

Peeter's model has been experimentally verified considering (i) the NB-DBD configuration where the β values issued from the {3} {9} equations system have been compared with those resulting from the volumes of non-ionized and ionized gases in the gap region, (ii) the PB-DBD configuration, where LCRmeter has demonstrated a strong correlation between the $\delta$ parameter and the gap capacitance, offering the ability to accurately assess seeds capacitance ($C_{seeds}$) as well as to validate the relation {7}

At the light of the experiments carried out, it is worth underlying the main performances and limitations of the equivalent electrical model:

(i) Seeds are considered as impenetrable pellets, i.e. plasma streamers cannot penetrate inside them through smaller pores,

(ii) Seeds must have a sufficiently elevated equivalent series resistance to be modelled by a single capacitor. Otherwise, a capacitor in parallel with a resistor will be a mandatory for modelling seeds with high water content,

(iii) Seeds are considered in a native and permanent anhydrous state with typical moisture content lower than 1%wt. If seeds are exposed to longer exposure times, their modelling (capacitor and/or resistor) should be considered as variable vs time to take into account the gradual loss of relative humidity within the seed,

(iv) The capacitance values obtained by the equivalent electrical model are very close to the ones measured by the LCRmeter since discrepancies remain in overall lower than 6%.

The important information provided by the equivalent electrical model must be emphasized:

(i) The capacitance of the seedbed can be measured during the operation of the discharge (which is not possible using a conventional LCRmeter). The variations of $C_{seeds}$ can be monitored as a function of treatment time. A seedbed that loses its water content during treatment could see its electrical parameters change (in particular capacitance and resistance) and the model will assess them.
(ii) In our PB-DBD configuration, contact points only appear when oxygen is added to helium. In fact, each contact point corresponds to a small volume occupied by micro-discharges. The determination of the β parameter permits to evaluate the total volume of these contact points and therefore to deduce additional parameters like energy density in contact points regions.
(iii) The electrical parameters (charge per filament, dimensionless parameters (α, β, γ, δ), breakdown voltages, …) can be distinguished on the negative half-cycles but also on the positive half-cycles. As an example, the model can highlight if total volume of contact points is different upon negative and positive half cycles.

In the case of the NB-DBD reactor, we have explained how $(Q\text{-}V)_{plasma}$ diagrams can be extracted from $(Q\text{-}V)_{reactor}$ diagrams to deduce relevant electrical parameters. The importance of distinguishing negative and positive values of these parameters has been underlined, in particular for $\beta^-$, $\beta^+$, $\xi_{OFF}^-$, $\xi_{OFF}^+$, $\xi_{ON}^-$ and $\xi_{ON}^+$. In some cases, the number of micro-discharges over a polarized half-cycle (lateral side of the Lissajous curve) is too low and does not allow the determination of electrical parameters. The admixture of oxygen with helium gas drives to a significant decrease in $\xi_{OFF}$ as well as to a substantial increase of $V_{br}^-$ and $V_{br}^+$ voltages and finally to a rise of the charge per filament only upon positive half-cycles ($Q_f^+$). In the case of the PB-DBD reactor, the equivalent electrical model has been implemented with γ and δ parameters to better understand how seeds can mitigate/enhance electrical properties of the discharge. Two groups of seeds have been distinguished among the six varieties of seeds activated in the PB-DBD. The differences of these groups rely on their respective capacitances ($C_{seeds}$) estimated by Lissajous diagrams with values close to 14.2 pF and 20.4 pF for groups I and II respectively.

Among the 6 types of seeds exposed to the same plasma treatment, 4 of them saw an improvement of their germinative properties, especially their vigor (through the $T_{50}$ parameter). The most significant effects were obtained on bean and lentil seeds: two agronomic specimens belonging to group II. Despite the close discrepancies of the $C_{seeds}$ values from group I and group II, one can reasonably assume that seeds with higher capacitance can receive higher dormancy release after plasma processing in PB-DBD. The reason is that they can locally enhance the electric field distribution and promote RO(N)S reactions at contact points. Those plasma-triggered biological effects should be even more pronounced comparing seeds with different water contents.

# Acknowledgements

This work has been achieved within the LABEX Plas@Par fundings and has been supported by grants from Région Ile-de-France (Sesame, Ref. 16016309) and Sorbonne Université Platform programme. This work is partly supported by French network GDR 2025 HAPPYBIO.

# Data availability

The data that support the findings of this study are available from the corresponding author upon reasonable request.

# References

[1] Asif Hussain Khoja, Muhammad Tahir, Nor Aishah Saidina Amin, Dry reforming of methane using different dielectric materials and DBD plasma reactor configurations, Energy Conversion and Management, 2017, Volume 144, 262-274

[2] A. M. Vandenbroucke, R. Morent, N. De Geyter, C. Leys, Non-thermal plasmas for non-catalytic and catalytic VOC abatement, Journal of Hazardous Materials 195 (2011) 30-54

[3] V. Demidyuk, J. C. Whitehead, Influence of Temperature on Gas-Phase Toluene Decomposition in Plasma-Catalytic System . Plasma Chem Plasma Process 27, 85–94 (2007)

[4] Zaka-ul-Islam Mujahid, Ahmed Hala, Plasma dynamics in a packed bed dielectric barrier discharge (DBD) operated in helium, Journal of Physics D: Applied Physics, Volume 51, Number 11, (2018), 11LT02, 1-6

[5] Y. Zhang, H.-Y. Wang, Wei Jiang, A. Bogaerts, Two-dimensional particle-in cell/Monte Carlo simulations of a packed-bed dielectric barrier discharge in air at atmospheric pressure, New J. Phys., 17 (2015) 083056

[6] A. Bogaerts, Q.-Z. Zhang, Y.-R. Zhang, K. Van Laer, W. Wang, Burning questions of plasma catalysis: Answers by modeling, Catalysis Today, 2019, Volume 337, Pages 3-14

[7] Ana Gómez-Ramírez, Antonio M. Montoro-Damas, José Cotrino, Richard M. Lambert, Agustín R. González-Elipe, About the enhancement of chemical yield during the atmospheric plasma synthesis of ammonia in a ferroelectric packed bed reactor, Plasma Process Polym. 2017;14:e1600081

[8] J. C. Whitehead, Plasma–catalysis: the known knowns, the known unknowns and the unknown unknowns, Journal of Physics D: Applied Physics, 2016, Volume 49, Number 24

[9] X. Zhang, C. S.-M. Lee, D. M. P. Mingos, D. O. Hayward, Oxidative coupling of methane using microwave dielectric heating, Applied Catalysis A: General 249 (2003) 151-164

[10] Y. Uytdenhouwen, S. Van Alphen, I. Michielsen, V. Meynen, P. Cool, A. Bogaerts, A packed-bed DBD micro plasma reactor for CO2 dissociation: Does size matter?, Chemical Engineering Journal, Volume 348, 2018, Pages 557-568

[11] Koen Van Laer, Annemie Bogaerts, Improving the Conversion and Energy Efficiency of Carbon Dioxide Splitting in a Zirconia-Packed Dielectric Barrier Discharge Reactor, Energy Technol. 2015, 3, 1038–1044

[12] Michielsen Inne, Uytdenhouw en Yannick, Pype Judith, Michielsen Bart, Mertens J., Reniers F., Meynen Vera, Bogaerts Annemie, CO2 dissociation in a packed-bed DBD reactor : first steps towards a better understanding of plasma catalysis, Chemical engineering journal, 326(2017), p. 477-488

[13] K. Takaki, J.-S. Chang, K. G. Kostov, Atmospheric pressure of nitrogen plasmas in a ferroelectric packed bed barrier discharge reactor. Part I. Modeling, IEEE Transactions on Dielectrics and Electrical Insulation, vol. 11, no. 3, pp. 481-490, 2004

[14] R. Wegst, M. Neiger, H. Russ, S. Liu, Experimental and theoretical investigations of removal of NOx from diesel-type engine exhaust using dielectric barrier discharges SAE Technical Paper, 1999, 1999-01-3686

[15] Fitzsimmons C., Shawcross J.T. and Whitehead J.C. 1999 Plasma-assisted synthesis of $N_2O_5$ from $NO_2$ in air at atmospheric pressure using a dielectric pellet bed reactor *J. Phys. D: Appl. Phys.* **32** 1136-1141.

[16] Fitzsimmons C., Ismail F., Whitehead J.C. and Wilman J.J 2000 The Chemistry of Dichloromethane Destruction in Atmospheric-Pressure Gas Streams by a Dielectric Packed-Bed Plasma Reactor *J. Phys. Chem. A* **104** 6032–6038.

[17] Cal M.P. and Schluep M. 2001 Destruction of Benzene in a Dielectric Barrier Discharge Plasma Reactor *Environ. Prog*. **20** 151-156.

[18] Jwa E., Lee S.B., Lee H.W., Mok Y.S. 2013 Plasma-assisted catalytic methanation of CO and $CO_2$ over Ni–zeolite catalysts *Fuel Processing Technol.* **108** 89-93.

[19] Zhou A., Chen D., Ma C., Yu F. and Dai B. 2018 DBD Plasma-ZrO2 Catalytic Decomposition of $CO_2$ at Low Temperatures *Catalysts* **8** 256

[20] D. Mei, X. Tu, Atmospheric Pressure Non-Thermal Plasma Activation of CO2 in a Packed-Bed Dielectric Barrier Discharge Reactor, ChemPhysChem, 2017, 18, 3253-3259

[21] K. Sacilik, A. Colak, Determination of dielectric properties of corn seeds from 1 to 100 MHz, Powder Technology 203 (2010) 365–370

[22] S. O. Nelson, S. Trabelsi, A century of grain and seed moisture measurement by sensing electrical properties, Transactions of the ASABE, 2012 American Society of Agricultural and Biological Engineers ISSN 2151-0032, Vol. 55(2): 629-636

[23] Meng Y., Qu G., Wang T., Sun Q., Liang D. and Hu S. 2017 Enhancement of Germination and Seedling Growth of Wheat Seed Using Dielectric Barrier Discharge Plasma with Various Gas Sources *Plasma Chem. Plasma P.* **37** 1105–1119.

[24] Puligundla P., Kim J.W. and Mok C. 2017 Effect of corona discharge plasma jet treatment on decontamination and sprouting of rapeseed (Brassica napus L.) seeds *Food Control* **71** 376-382

[25] Filatova I.I., Azharonok V.V., Goncharik S.V., Lushkevich A., Zhukovsky A.G. and Gadzhieva G.I., 2014 Effect of RF plasma treatment on the germination and phytosanitary state of seeds, *J. Appl. Spectrosc.,* Vol 81, Number 2, 250-256

[26] Šerá B., Stranák V., Serý M., Tichý M. and Spatenka P. 2008 Germination of Chenopodium Album in Response to Microwave Plasma Treatment *Plasma Sci. Technol.* **10** 506-511.

[27] W. A. Pryor, Oxy-radicals and related species: their formation, lifetimes, and reactions, Ann. Rev. Physiol. 1986. 48:657-67 Copyright © 1986 by Annual Reviews

[28] N. Puac, M. Gherardi, M. Shiratani, Plasma agriculture: a rapidly emerging field, Plasma Processes and Polymers, 2018, Volume 15, Issue 2, e1700174, 1-5

[29] B. Kucera, M. Alan Cohn, G. Leubner-Metzger, Plant hormone interactions during seed dormancy release and germination, Seed Sci. Res., 2005, 15, 281–307

[30] R. Zukiene, Z. Nauciene, I. Januskaitiene, G. Pauzaite, V. Mildaziene, K. Koga, M. Shiratani, Dielectric barrier discharge plasma treatment-induced changes in sunflower seed germination, phytohormone balance, and seedling growth, Applied Physics Express, 2019, Volume 12, Number 12

[31] L. Wojtyla, K. Lechowska, S. Kubala, M. Garnczarska, Different Modes of Hydrogen Peroxide Action During Seed Germination, Front. Plant Sci., 2016, 7, 66

[32] S. Sadhu, R. Thirumdas, R. R. Deshmukh, U. S. Annapure, Influence of cold plasma on the enzymatic activity in germinating mung beans (Vigna radiate), LWT, 2017, Volume 78, 97-104

[33] A. Iranbakhsh, Z. O. Ardebili, H. Molaei, N. O. Ardebili, M. Amini, Cold Plasma Up-Regulated Expressions of WRKY1 Transcription Factor and Genes Involved in Biosynthesis of Cannabinoids in Hemp (Cannabis sativa L.), Plasma Chemistry and Plasma Processing, Volume 40, 527-537 (2020)

[34] Sera B. and Sery M. 2018 Non-thermal plasma treatment as a new biotechnology in relation to seeds, dry fruits, and grains *Plasma Sci. Technol.* **20** 044012

[35] Bormashenko E., Grynyov R., Bormashenko Y. and Drori E. 2012 Cold Radiofrequency Plasma Treatment Modifies Wettability and Germination Speed of Plant Seeds *Sci. Rep.* **2** 741.

[36] Schnabel U., Niquet R., Krohmann U., Winter J., Schlüter O., Weltmann K.D. and Ehlbeck J. 2012 Decontamination of Microbiologically Contaminated Specimen by Direct and Indirect Plasma Treatment *Plasma Process Polym.* **9** 569-575

[37] D. Butscher, Hanne Van Loon, Alexandra Waskow, Philipp Rudolf von Rohr, M. Schuppler, Plasma inactivation of microorganisms on sprout seeds in a dielectric barrier discharge, International Journal of Food Microbiology, 2016, Volume 238, 222-232

[38] P. Puligundla, J.-W. Kim, C. Mok, Effect of corona discharge plasma jet treatment on decontamination and sprouting of

rapeseed (Brassica napus L.) seeds, Food Control, 2017, Volume 71, 376-382

[39] Butterworth T. and Allen R. 2016 Plasma-catalyst interaction studied in a single pellet DBD reactor: dielectric constant effect on plasma dynamics *Plasma Sources Sci. Technol.* **26** 065008.

[40] Brandenburg R. 2017 Dielectric barrier discharges: progress on plasma sources and on the understanding of regimes and single filaments *Plasma sources sci. technol.* **26** 053001.

[41] Kostov K.G., Honda R.Y., Alves L.M.S. and Kayama M.E. 2009 Characteristics of Dielectric Barrier Discharge Reactor for Material Treatment *Braz. J. Phys.* **39** 322-325.

[42] Peeters F., *The electrical dynamics of dielectric barrier discharges*, PhD Thesis, Eindhoven: Technische Universiteit Eindhoven, 2015.

[43] Dielectric Properties of Grain and Seed in the 1 to 50-MC Range, S. O. Nelson, Transactions of the ASAE, Vol. 8, No. 1, pp 38-48, 1965

[44] Measurement of Electrical Properties of Rapeseed Seeds with LCR Meter Good Will8211, K. Kardjilova, E. Bekov, Z. Hlavacova, A. Kertezs, International Journal of Applied Science and Technology, Vol. 2 No. 8; October 2012

[45] I. Biganzoli, R. Barni, A. Gurioli, R. Pertile, C. Riccardi, Experimental investigation of Lissajous figure shapesin planar and surface dielectric barrier discharges, Journal of Physics: Conference Series550 (2014) 012039

[46] Feng Liu, Weiwei Wang, Xijiang Chang, Zhonghang Wu, Long He, Zebin Li, Zhijiang He, Rongqing Liang, Electrical characteristics of monofilaments in dielectric barrier discharge plasma jets at atmospheric pressure, EPL, 97 (2012) 65001

[47] T.Butterworth, R.Elder, R.Allen, Effects of particle size on CO2 reduction and discharge characteristics in a packed bed plasma reactor, Chemical Engineering Journal, Volume 293, 1 June 2016, Pages 55-67

[48] Mangina R., Enloe C. and Font G. 2015 Influence of oxygen content on the characteristics of DBD plasma actuator *J. Phys.: Conf. Ser.* **635** 042009.

[49] Erfani R., Zare-Behtash H., Hale C. and Kontis K. 2015 Development of DBD plasma actuators: The double encapsulated electrode *Acta astronaut.* **109** 132-143.

[50] Bian D.-L., Wu Y., Jia M., Long C.-B. and Jiao S.-B. 2017 Comparison between AlN and Al2O3 ceramics applied to barrier dielectric of plasma actuator *Chinese Phys. B* **26** 084703

[51] Joussot R., Leroy A., Weber R., Rabat H., Loyer S. and Hong D. 2013 Plasma morphology and induced airflow characterization of a DBD actuator with serrated electrode *J. Phys. D: Appl. Phys* **46** 125204.